**Review article**

# Atomically flat single terminated oxide substrate surfaces

Abhijit Biswas[a], Chan–Ho Yang[b], Ramamoorthy Ramesh[c,d,e], and Yoon H. Jeong[a*]

[a] Department of Physics, Pohang University of Science and Technology, Pohang 790-784, South Korea

[b] Department of Physics and Institute for the NanoCentury, KAIST, Daejeon 305-701, South Korea

[c] Department of Materials Science and Engineering, University of California, Berkeley, CA 94720, USA

[d] Department of Physics, University of California, Berkeley, CA 94720, USA

[e] Materials Science Division, Lawrence Berkeley National Laboratory, Berkeley, CA 94720, USA

***Corresponding author**.

E-mail addresses: 01abhijit@gmail.com (Abhijit Biswas), chyang@kaist.ac.kr (Chan-Ho Yang), rramesh@berkeley.edu (Ramamoorthy Ramesh), yhj@postech.ac.kr (Yoon H. Jeong)

## ABSTRACT

Scientific interest in atomically controlled layer-by-layer fabrication of transition metal oxide thin films and heterostructures has increased intensely in recent decades for basic physics reasons as well as for technological applications. This trend has to do, in part, with the coming post-Moore era, and functional oxide electronics could be regarded as a viable alternative for the current semiconductor electronics. Furthermore, the interface of transition metal oxides is exposing many new emergent phenomena and is increasingly becoming a playground for testing new ideas in condensed matter physics. To achieve high quality epitaxial thin films and heterostructures of transition metal oxides with atomically controlled interfaces, one critical requirement is the use of atomically flat single terminated oxide substrates since the atomic arrangements and the reaction chemistry of the topmost surface layer of substrates determine the growth and consequent properties of the overlying films. Achieving the atomically flat and chemically single terminated surface state of commercially available substrates, however, requires judicious efforts because the surface of as-received substrates is of chemically mixed nature and also often polar. In this review, we summarize the surface treatment procedures to accomplish atomically flat surfaces with single terminating layer for various metal oxide substrates. We particularly focus on the substrates with lattice constant ranging from 4.00 Å to 3.70 Å, as the lattice constant of most perovskite materials falls into this range. For materials outside the range, one can utilize the substrates to induce compressive or tensile strain on the films and explore new states not available in bulk. The substrates covered in this review, which have been chosen with commercial availability and, most importantly, experimental practicality as a criterion, are $KTaO_3$, $\mathit{RE}ScO_3$ (*RE* = Rare-earth elements), $SrTiO_3$, $La_{0.18}Sr_{0.82}Al_{0.59}Ta_{0.41}O_3$ (LSAT), $NdGaO_3$, $LaAlO_3$, $SrLaAlO_4$, and $YAlO_3$. Analyzing all the established procedures, we conclude that atomically flat surfaces with selective A- or B-site single termination would be obtained for most commercially available oxide substrates. We further note that this topmost surface layer selectivity would provide an additional degree of freedom in searching for unforeseen emergent phenomena and functional applications in epitaxial oxide thin films and heterostructures with atomically controlled interfaces.

**Contents**

## 1. Introduction

Research in the field of transition metal oxides (TMOs) have shown tremendous progress for the last few decades in condensed matter physics and materials science communities not only because of its richness in fundamental science but also due to its tremendous application potential **[1-3]**. TMOs exhibit remarkably diverse functional properties such as ferroelectricity, magnetism, high temperature superconductivity (HTSC), colossal magnetoresistance (CMR), metal-insulator transitions (MIT), multiferroicity (coexistence of ferroelectricity and magnetism), anomalous Hall effect (AHE), topological insulator (TI), and quantum spin-liquid (QSL) [**4-9**]. These functional properties of TMOs can be exploited for practical applications such as spintronic (spin-based electronics) devices, resistance random access memory (ReRAM), solid oxide fuel cells (SOFCs), light-emitting diodes (LEDs), transparent conductors, and solar cells **[4-14]**. In post-Moore era, in particular, electronic devices with multifunctionality may offer a new alternative to replace the current silicon-based technology because the additional value the devices would generate from multifunctionality may create an economically viable path superseding the miniaturization limit of silicon electronic devices **[15-19]**. In this perspective, oxide electronics based on multifunctional properties of TMOs looks indeed promising. Among various TMOs, the case of more familiar perovskites would make a point. Perovskites have traditionally received ample attention because they exhibit all of the functional properties mentioned above in a particularly simple structure **[20-24]**.

Perovskites are a class of materials of $ABO_3$ type, in which each A-site cation is surrounded by twelve oxygen anions and each B-site cation of smaller size is surrounded by six oxygen anions that form a $BO_6$ octahedron. Ideal perovskite $ABO_3$ has a cubic structure with a lattice constant of ~4.00 Å **[24-27]**. Most of the electronic properties of perovskites are determined by the physics associated with the transition metal and the corner-sharing oxygen anions of the $BO_6$ octahedra. Depending upon the radius and position of the B-site cation, the cubic structure becomes distorted due to the tilting and distortion of the $BO_6$ octahedra. According to Goldschmidt's rule **[28]**, deformation occurs when the relative size of the A-site cation compared to the B-site cation becomes too small; the oxygen anions move towards the central A-site to shrink the empty volume around the A-site cation and consequent tilting of the $BO_6$ octahedra follows. This overall shrinkage deforms the unit cell leading to the reduction of the cubic symmetry, and the structure changes to relatively low symmetry ones, e.g., orthorhombic, rhombohedral, tetragonal, hexagonal. The beauty of perovskite $ABO_3$ structure is that almost half of the elements in the periodic table can be accommodated into the structure at either the A-site or the B-site. Also, even under ambient conditions (e.g., pressure and temperature) the B-site cations can assume different valences and coordination numbers and this fact gives rise to a virtually unlimited number of $ABO_3$ perovskites **[24-27]**. Thus, the elemental and structural diversities occurring in perovskites provide enormous opportunities to explore a variety of multifunctional properties based on the structure-property relationship. Of course, perovskites are a representative example and similar opportunities would also exist in other TMOs with diverse structures of rock-salt, corundum, rutile, spinel, garnet, etc.

Even more exciting, however, is the fact that advanced thin film growth techniques with atomic controllability provide further opportunities to design and synthesize artificial complex TMO heterostructures and superlattices to bring forth emergent physical properties, normally not seen in bulk states [**29-49**]. For example, one of the breakthroughs in TMO thin film research came with the observation of high mobility two-dimensional electron gas (2DEG) at the interface between two non-conducting oxides and subsequent discovery of associated interface phenomena such as superconductivity, magnetism, magneto-electric coupling, and quantum Hall effect [**50-60**]. In this era of nanoscience and nanotechnology, various techniques to grow TMO thin films have been developed rapidly since 1980s **[61]**. Pulsed laser deposition (PLD), for instance, has been a technique frequently adopted for precise thin film growth of TMO materials, because it is most forgiving in situations involving multiple metal ions and oxygen gas. This technique as well as others such as molecular beam epitaxy (MBE) or magnetic sputtering technique can control the deposition process down to atomic scale and thus have been used to design and synthesize various artificial TMO heterostructures **[62-66]**. Generally speaking, thin film growth methods can provide several advantages over bulk ones in synthesizing materials; in particular, they are capable of producing metastable phases of various materials that would not exist in bulk form. In epitaxial TMO films, strain can play a role of an extra degree of freedom to expand the parameter space for tuning material properties. Most dramatically, thin film growth methods can be used to achieve atomically controlled heterostructures, multilayers, and superlattices of constituent oxides.

In order to fabricate functional TMOs in ultra-thin film or artificial structural form, chemically and structurally compatible underlying substrates are essential because the interface interaction between an overlying thin film and the underlying substrate surface critically controls the growth and consequent properties of TMO thin films **[35,36,45,48,65,67]**. In other words, substrates with atomically flat and chemically homogeneous surfaces, i.e., surfaces with the step-terrace structure of *one*-unit cell step height and also of *single* chemical termination, are indispensable for the growth of high quality epitaxial thin films with atomically controlled interface. It is noteworthy that metal oxide perovskites are also of importance as substrates because many single crystals available as substrates are of perovskite or perovskite-related structures. Commercially accessible crystals, for example, include $KTaO_3$, $\textit{RE}ScO_3$ (*RE* = Dy, Gd, and Nd), $SrTiO_3$, $La_{0.18}Sr_{0.82}Al_{0.59}Ta_{0.41}O_3$ (LSAT), $NdGaO_3$, $LaAlO_3$, $SrLaAlO_4$, and $YAlO_3$. (Table **I**) **[68]**. The lattice constant of substrates as well as multifunctional perovskites of scientific interest varies widely from 4.00 Å to 3.70 Å as shown in Fig. **1**. The diversity of the substrates as evident from the figure means that for a given material a lattice-matched substrate can be chosen to produce a film in its most natural state. It also means that either compressive or tensile strain can be induced in the film by choosing appropriate lattice-mismatched substrates to alter the functionalities. However, to fully take advantage of these widely varying substrates in growing thin films or heterostructures of complex functional oxides, one would like to establish the procedures for treating various as-received substrates to secure the substrate surface to be atomically flat with step-terrace structures of one-unit cell step height and chemically homogeneity. It is noted that the surface of as-received perovskite substrates is usually of chemically mixed nature with coexisting A-site cations and B-site cations. More often than not the surface of metal oxide crystals is polar, and thus thermodynamic as well as electrostatic aspects must be taken into consideration to account for the stability of substrate surfaces. In addition, the mixed chemical nature may be the simple consequence of a difficulty of cutting a single crystal along a particular crystallographic direction and subsequent mechanically polishing of the surface. These mechanical procedures often end up with the topmost surface layer neither atomically flat nor chemically homogeneous.

**Table 1.** List of commercially available single crystal metal oxide substrates with preferable orientations used for thin film growth. The lattice constant ranges between 4.00 Å and 3.70 Å. Other substrates not listed here would also be available [**68**].

| **Substrate** | **Orientation** | **Structure** | **Lattice constants (Å)** | **Cubic (pseudo) lattice constant (Å)** |
|---|---|---|---|---|
| $NdScO_3$ | (110) | Orthorhombic | $a$ = 5.57<br>$b$ = 5.77<br>$c$ = 7.99 | 4.00 |
| $KTaO_3$ | (001) | Cubic | $a$ = 3.988 | 3.988 |
| $GdScO_3$ | (110) | Orthorhombic | $a$ = 5.48<br>$b$ = 5.76<br>$c$ = 7.92 | 3.96 |
| $DyScO_3$ | (110) | Orthorhombic | $a$ = 5.54<br>$b$ = 5.71<br>$c$ = 7.89 | 3.94 |
| $SrTiO_3$ | (001), (110), (111) | Cubic | $a$ = 3.905 | 3.905 |
| $La_{0.18}Sr_{0.82}$ $Al_{0.59}Ta_{0.41}O_3$ (LSAT) | (001) | Cubic | $a$ = 3.88 | 3.88 |
| $NdGaO_3$ | (001), (110) | Orthorhombic | $a$ = 5.43<br>$b$ = 5.50<br>$c$ = 7.71 | 3.86 |
| $LaAlO_3$ | (001) | Rhombohedral | $a$ = 3.78 | 3.78 |
| $SrLaAlO_4$ | (001), (100) | Tetragonal | $a$ = 3.75<br>$c$ = 12.63 | 3.75 |
| $YAlO_3$ | (110) | Orthorhombic | $a$ = 5.18<br>$b$ = 5.33<br>$c$ = 7.37 | 3.72 |

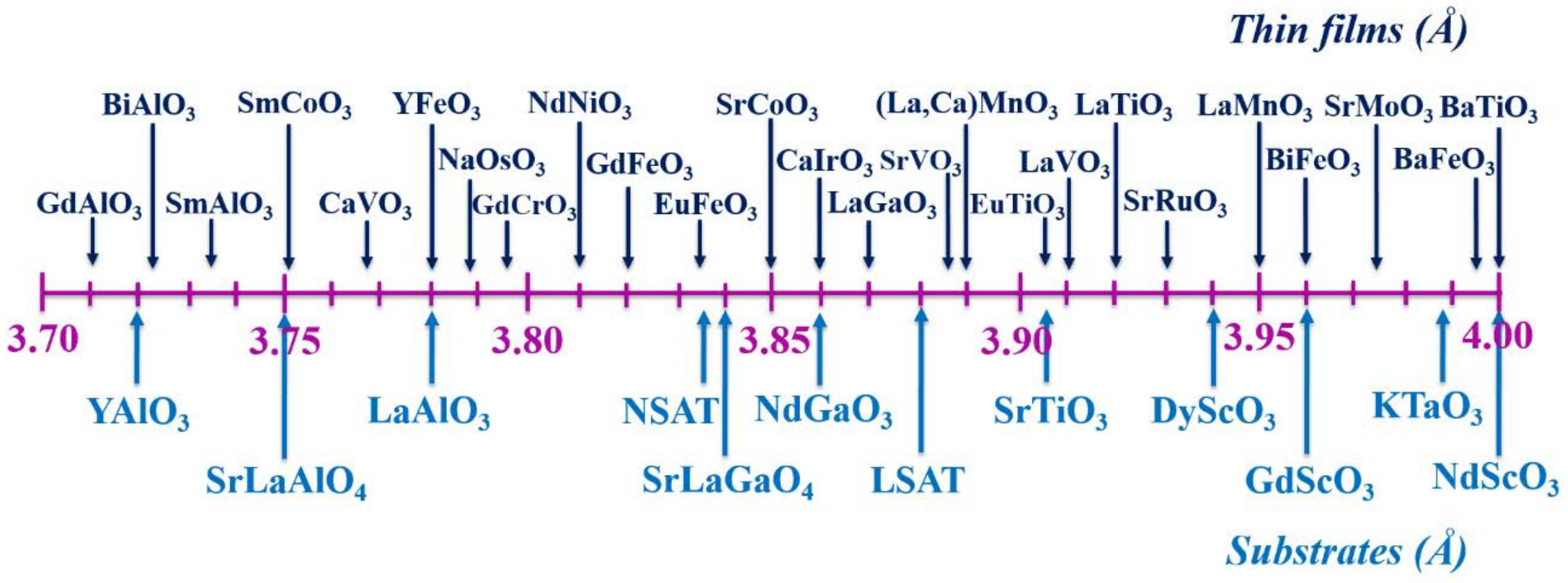

**Fig. 1**. (Color online) Comparative lattice constant (Å) of various perovskite thin films and commercially available substrates within the range 4.00 Å ~ 3.70 Å. Cubic (or pseudo-cubic) lattice constants are taken from the literature **[25-27]**.

Considering the progress and proliferation of thin film deposition of functional TMOs in recent years, it appears to be of value and timely to the thin film community to collect and summarize the surface treatment methods and characterization procedures for commercially available substrates. In fact, various methods are available in the literature depending on the chemical stability, imperfection of the crystal, defects, and miscut angle of the substrates. In this review, we will focus on how to turn the surfaces of various as-received substrates with chemically mixed termination to atomically flat single terminated surfaces with step-terrace structures of *one*-unit cell step height; in addition, we also consider the methods of identifying the surface termination of the treated substrates. The foremost goal of the present manuscript is to provide a practical guidance for choosing and preparing substrates for the deposition of complex TMO thin films or heterostructures and also to furnish information about topmost surface termination for oxide substrates with lattice constant in the range from 4.00 Å to 3.70 Å. Fundamentally, substrate topmost layer termination not only plays a crucial role in obtaining exotic interfacial phenomena but also helps to interpret various emerging non-bulk-like phenomena observed in atomically controlled thin films or superlattices **[69]**. It is to be mentioned here that single terminated substrate surfaces are also of critical importance from surface science point of view; for example, the phenomena such as site-specific adsorption and catalytic reactions would strongly depend on the topmost layer arrangement and its chemistry **[70,71]**. Even for theoretical calculations, substrate surface termination is considered an important parameter for explaining interfacial phenomena as well as for designing new type of interfaces based on various combinations of constituent layers **[72-75]**.

# 2. Principle of surface treatments and surface characterizations

## 2.1. Chemistry of surface treatments

Commercially available substrate crystals usually come with mechanically polished surfaces of low miscut angle, typically less than 0.1°. (High miscut substrates are also available separately.) Miscut angle, of course, is the angle between the normal to the substrate surface and to the crystallographic plane. Along the [001] direction, a perovskite $ABO_3$ crystal consists of alternate stacking of AO and $BO_2$ layers, and the surface of as-received substrates is always composed of a mixed layer of AO and $BO_2$. Similar features of mixed termination with different chemical compositions are found for substrates with other (*hkl*) orientations. Depending upon the oxidation state of metallic elements, the ideal topmost surface layer would be either polar or non-polar. Over the last few decade extensive efforts have been spent to study and understand the physics and chemistry of metal oxide surfaces, especially the polar ones [**76-79**]. Considering that electrostatics tends to require charge neutrality even on polar surfaces and thus associated surface reconstructions would follow, it is extremely challenging to remove a particular type of cations from a polar surface to obtain single termination that would come necessarily with compensating surface charges. Although the microscopic processes of reconstruction and compensation are complex for polar surfaces, the final compensated surface would be reached either by electronic reconstruction via partial filling of electronic surface states or by changes in the surface stoichiometry (or atomic reconstruction) via spontaneous desorption of atoms, faceting, ordering of surface vacancies, etc. or by combination of electronic and atomic reconstructions **[77-79]**.

As for the polar surfaces of metal oxides in general, it may be stated that the precise lowest energy state atomic arrangements, the exact electronic structures, and the stoichiometry of many surfaces still remain somewhat unresolved at the present time. Irrespective of intrinsic microscopic reconstructions of metal oxide surfaces, however, one needs to obtain atomically flat surfaces with single termination from substrate crystals for controlled epitaxial thin film growth. In this regard, various empirical efforts have been successful, at least for thin film growth purposes, in establishing the treatment procedures to achieve atomically flat and chemically homogeneous surfaces for many metal oxide substrates. The procedures would involve one or some combination of the followings: (1) only thermal annealing, (2) wet acidic solution etching, (3) wet basic solution etching, (4) deionized-water leaching and thermal annealing, and (5) thermal annealing in cation-rich environments. Thermal annealing often leads to a segregation of metallic oxides and an enrichment of a particular cation at the surface; it may also produce atomically flat surfaces. Obviously, electrostatic energy plays an important role in the segregation process on the surface. To achieve a surface that has only one type of termination (either A- or B-type), chemical treatments often play an essential role because one particular chemical species on the surface can be selectively removed using a wet chemical etching process, either acidic or basic, judiciously chosen for a given substrate and a given orientation. The selectivity for the (001) orientation, for example, is due to the fact that AO and $BO_2$ residing on the crystal surface generally differ in their chemical properties, particularly in the solubility of their hydroxyl groups in acids or water.

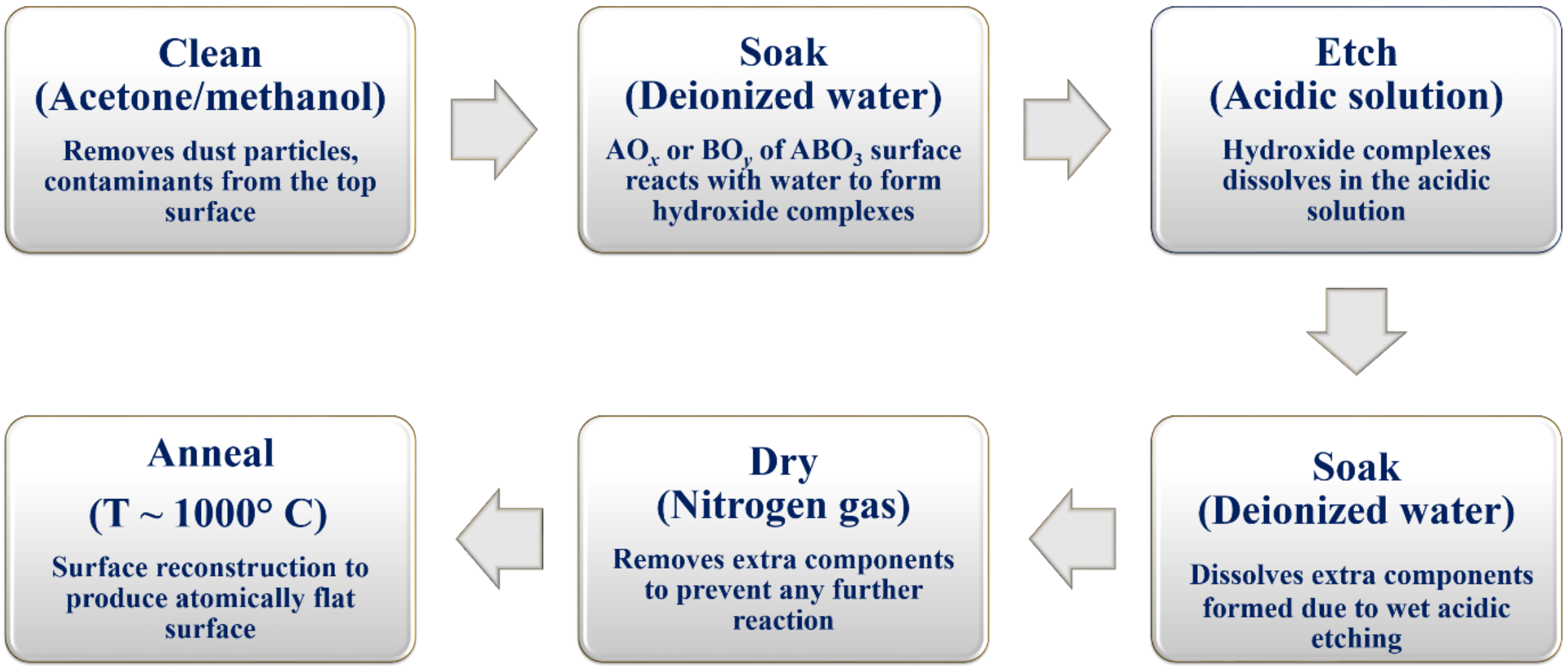

**Fig. 2.** (Color Online) Schematic procedures for producing atomically flat and chemically single terminated surfaces of metal oxide substrates. Wet acidic surface treatment is adopted most often.

The first successful preparation of an atomically flat surface with chemical homogeneity was carried out by M. Kawasaki *et al.* for the most popular substrate in TMO thin film growth, $SrTiO_3$ (001) **[80]**. They pioneered an acidic etching method relying on the acidity/basicity difference of different metal oxides on the surface, and this method still remains one of the most adopted ones in preparing the smooth surface of many metal oxide substrates. The typical procedures of the wet acidic etching method to remove one type of metal oxide preferentially are summarized in **Fig. 2**: (1) Substrates are cleaned with acetone and methanol in an ultrasonic bath for 10-20 min each to remove dust particles. (2) The substrates are soaked in deionized (DI) water for 5-10 min in an ultrasonic bath, and the DI-water soaking transforms differentially either AO or $BO_2$ on the surface to a hydroxide complex, depending on the relative chemical reactivity of the species. (3) The hydroxide complexes on the surface are removed, using buffered hydrofluoric acid (BHF) of $NH_4F$: HF = 7:1 with pH = 4.5-5.5 for 30 sec to a few minute treatments. (BHF is also called buffered oxide etchant or BOE.) The chemically treated substrates are then annealed for 2 to 3 hours at a temperature around ~1000 °C to obtain atomically flat surfaces with step-terrace structures of one unit-cell step height. This thermal treatment typically reconstructs the surface and produces an atomically flat layer with chemically homogeneous single termination. In some cases, optionally another DI-water rinsing may be applied to obtain fully single terminated atomically flat substrates. In applying the wet etching method, other etching solutions such as $HCl+HNO_3$, $HCl+NH_4OH$, or NaOH may also be utilized depending on the nature of chemical reactivity and electrostatic bonding of the metallic elements on the surface with surrounding oxygen atoms.

Deionized water leaching was also adopted as an alternative method to produce atomically flat surfaces **[81]**; in this method, substrates are annealed first at a high temperature to induce a segregation of selective type of cations on the surface, and these cations are leached away via deionized water soaking. Another heat treatment method with a reverse idea was also devised for substrate treatments [**82**]; in this method, cations are supplied, instead of being etched away, from outside to compensate for the loss of cations from the substrate by high temperature annealing (cation rich treatment). By using the method, atomically flat surfaces with step-terrace structures of one-unit cell step height were successfully obtained for certain substrates. Details about the cation rich treatments are discussed in Section **3.3** along with other surface treatment methods.

### 2.2. Surface characterizations

After treating substrates according to the surface treatment procedures, the surface must be characterized to ensure structural flatness and chemical homogeneity. For this purpose, two kinds of surface measurements are generally conducted: atomic scale structural topography to check the flatness and chemical analysis to confirm the topmost layer termination. Structural characterization of the surface is rather straightforward and atomic force microscopy (AFM) is usually used. For instance, the surface of an as-received $ABO_3$ (001) substrate would be far from being flat and homogeneous (**Fig. 3a**), and this can be seen easily from the corresponding AFM image (**Fig. 3b**). After the prescribed treatment procedures, the substrate surface turns into a step-terrace structure (Fig. **3c**) and the AFM image is able to show it clearly (**Fig. 3d**). For simple $ABO_3$ (001) substrates, the properly treated surfaces would be ideally terminated by either AO or $BO_2$ layer; however, in reality annealing at high temperatures always results in the loss of some oxygen's from the surface.

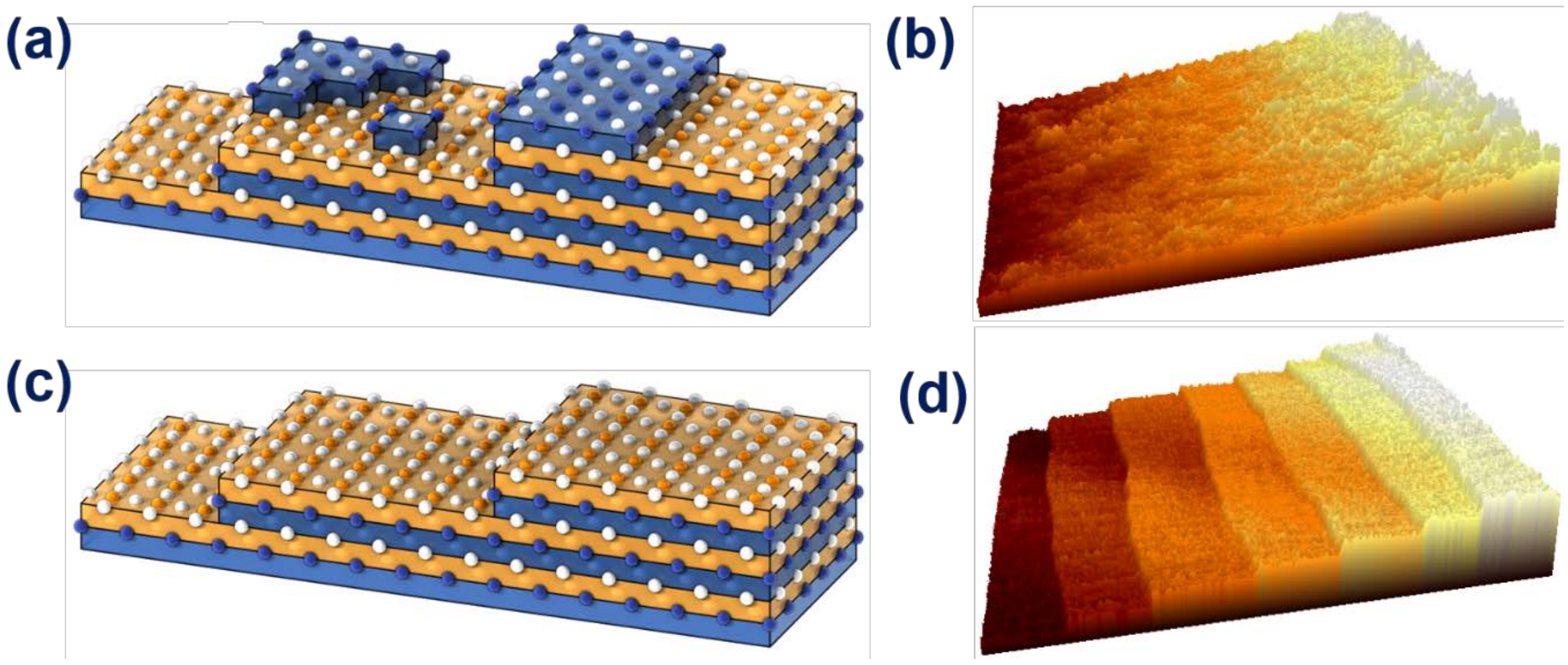

**Fig. 3.** (Color online) Atomic structure and typical AFM topography of: (a)-(b) an as-received surface and (c)-(d) an atomically flat surface. Schematic figures (a) and (c) are reproduced with permission from ref. **[168]**. Copyright © 2010 WILEY-VCH Verlag GmbH & Co. KGaA, Weinheim.

Since the knowledge whether the surface is terminated by A- or B-sites is critical in the successful growth of atomically controlled thin films, chemical characterization measurements are essential to confirm the topmost layer termination of the atomically flat surface. Some of the techniques previously adopted to reveal the chemical nature as well as the surface reconstruction include: ion scattering [**83,84**], lateral force microscopy (LFM) [**85**], scanning tunneling microscope (STM) [**86**], reflection high energy electron diffraction (RHEED) **[87]**, and various spectroscopic techniques such as low energy electron diffraction (LEED) [**88**], X-ray absorption spectroscopy (XAS) [**89**], and X-ray photoelectron spectroscopy (XPS) **[90]**. These techniques would provide the information about the richness of a particular type of cation on the surface. For instance, LFM would show a uniform frictional response when a substrate is singly terminated while STM and other spectroscopic methods are able to provide the information about various surface reconstructions and their atomic structures. RHEED is also frequently used to check the surface quality as uniform atomically flat surfaces would show sharp and narrow diffraction spots as well as Kikuchi lines, characteristics of crystalline surfaces. Ion scattering measurements are very effective and indispensable to thin film growers for substrate characterization, as it is able to provide the direct elemental information and its coverage about the topmost layer of metal oxide substrates. Among various ion scattering methods, special mention should be given to the time-of-flight mass spectrometry of recoiled ions (TOF-MSRI) because it has recently been in wide use to characterize the topmost layer of substrates and, in fact, many TOF-MSRI results are cited in this review. In TOF-MSRI, potassium ions with a relatively low kinetic energy (~10 keV) are typically used; $K^+$ ions are projected at sample surfaces at a low grazing angle for selectively probing first a few layers. The intensities of recoiled ions are strongly influenced by the ions that surround them on the surface as a result of shadowing and blocking effects, and thus vary with the incident azimuthal angle reflecting the crystalline symmetry of the surface. Thus, comprehensive angle resolved measurements would provide detailed information about the topmost layer of metal oxide substrates. The details on TOF-MSRI can be found elsewhere **[84]**.

## 3. Treated surfaces of various metal oxide substrates

To synthesize epitaxial thin films and heterostructures of functional oxides with precise atomic control at the interface between constituent layers, the first step is to secure atomically flat and chemically single terminated substrates with step-terrace structures of one-unit cell step height as stated in the previous sections. Obtaining atomically flat single terminated surfaces, however, with one-unit cell step height from commercially available substrates is a challenging task and would constitute an initial obstacle to be overcome by the researchers working with TMO thin films. In this main part of the review, we collect the successful concrete cases for surface treatments of popular metal oxide substrates and discuss the actual procedures of the surface treatment methods in detail.

### 3.1. As-received substrate surface

As described in the previous section, the basic structure of the $ABO_3$ perovskites with (001) orientation consists of alternating stacking of AO and $BO_2$ atomic layers which are electrostatically polar in many cases. The surface of commercially available metal oxide substrates in as-received state is usually composed of a mixture of AO and $BO_2$ layers, and does not exhibit the step-terrace structure of one-unit cell step height that is expected for single terminated atomically flat surfaces. For substrates with other (*hkl*) orientations, different combinations of metal oxide layers are exposed on the surface, but they also do not show step-terrace structures. Indeed, the AFM image of the surface of an as-received substrate shown in **Fig. 3b** confirm the lack of the step-terrace structure on the surface. The root-mean square surface roughness of as-received substrates are typically ≥ 1 nm. In the following sections, the details of the surface treatment of treated substrates are presented.

### 3.2. Wet etching and/or thermal annealing treatments

Substrate surface treatment using a chemical solution (acidic or basic) as an etchant and/or high temperature thermal annealing is one of the common methods adopted in the thin film laboratories. In this section, we review the cases for which this method has been used to achieve atomically flat one-unit cell step terrace structure, starting with the standard example of $SrTiO_3$ substrates.

#### 3.2.1. $SrTiO_3$ (001), (110), and (111) surfaces

$SrTiO_3$ is a chemically and compositionally stable cubic perovskite with an indirect band gap of 3.25 eV and a direct gap of 3.75 eV **[91]**. It has a very large dielectric constant of ~300 and resistivity of over $10^9$ Ω-cm at room temperature **[68]**. Also, high quality single crystals of $SrTiO_3$ are commercially available with different crystalline orientations, i.e., (001), (110), and (111). Thus, $SrTiO_3$ crystals have been used as substrates for growing almost all TMO thin films, including high $T_c$ superconductors, CMR materials, various ferroelectric and magnetic materials, and more specially for creating 2DEG at the film-substrate interface and various unexpected emergent properties in heterostructures grown on them **[50,92,93]**. $SrTiO_3$ surface is also of critical importance for observing 2DEG and electron liquid **[94,95]**. As a whole, $SrTiO_3$ crystals are probably the most important substrates for TMOs and, naturally, obtaining atomically flat and chemically controlled surfaces for $SrTiO_3$ substrates has been of prime importance to thin film growers. Thus, we allocate a sizable amount of space for the description of $SrTiO_3$ substrates. We note here for reference purposes that Sánchez *et al.* recently reviewed on the surface treatment of $SrTiO_3$ substrates **[96]**.

#### 3.2.1.1. $SrTiO_3$ (001) surface

The recipe for the surface treatments of $SrTiO_3$ (001) substrates was first developed by Kawasaki *et al.* **[80,97]** and was further refined later by Koster *et al.* **[98,99]**. $SrTiO_3$ (001) consists of stacked alternating layers of charge-neutral SrO and $TiO_2$ as depicted in **Fig. 4a** with a step height of 3.905 Å. The “Kawasaki” method relies on chemical treatments to distinguish SrO and $TiO_2$ chemically and make the surface chemically homogeneous. The method adopts a buffered hydrofluoric acid (BHF) etching solution with pH = 4.5 for the (001) surface. Strontium oxide SrO reacts more strongly in the etching solution than does $TiO_2$, and forms strontium hydroxide $Sr(OH)_2$ which dissolves easily. After etching for 30 s, the substrates are annealed at 1000 °C in $O_2$ gas flow for 2~3 h to get a step terrace structure; the annealing temperature is nominally kept around 1000°C to avoid Sr diffusion from the bulk to surface, which would occur at higher temperatures. The combination of wet etching and thermal annealing leads to atomic reorganization on the surface, and the surface becomes atomically flat and exhibits clear step-terrace structures with a sharp straight line at step edges with one-unit cell height as shown in **Fig. 4b [80]**. The (001) surface after the treatments is predominantly of $TiO_2$ as Sr related oxides are etched way from the surface; indeed, low energy ion scattering spectroscopy (ISS) measurements confirmed the Ti-rich termination as shown in **Fig. 4c [80]**.

High resolution synchrotron-radiation photoemission spectroscopy also showed that $TiO_2$ termination is more stable than the SrO termination, and re-etching the substrate after the high temperature annealing in an $O_2$ environment produces nearly perfect atomically flat surface **[100]**. Ohnishi *et al.* also reported that re-etching the substrate after annealing removes SrO coverage and produces the perfect $TiO_2$ terminated surface, as proved by coaxial impact-collision iron scattering spectroscopy (CAICISS) **[101]**. Frangeto *et al.* showed by grazing incidence x-ray diffraction that the chemically treated surface is ideally composed of ~75% $TiO_2$ and ~25% SrO **[102]**. By using the combination of AFM, LFM, XPS, and RHEED, Gunnarsson *et al.* confirmed the B-site termination for the atomically flat (001) $SrTiO_3$ surface **[103]**. We also were able to reproduce the atomically flat surface and confirmed the Ti-rich topmost surface layer using TOF-MSRI **[104]**. Raisch *et al.* used x-ray photoelectron diffraction (XPD) experiments to show $TiO_2$ surface termination (with negligible amount of SrO termination) **[105]**.

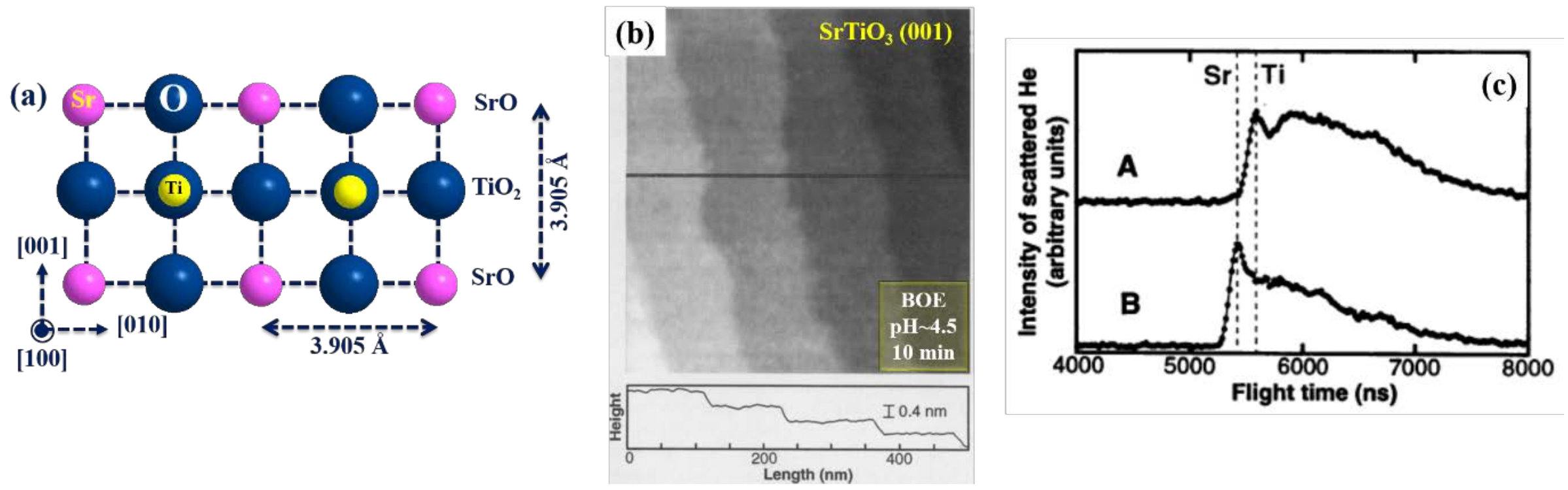

**Fig. 4.** (Color online) (a) Schematic side view of $SrTiO_3$ (001) surface showing alternating neutral SrO and $TiO_2$ layers with one-unit cell height of 3.905 Å. (b) Atomically flat surface with one-unit cell step-terrace structure was obtained by the wet acidic etching method. AFM image is shown. (c) Low energy ion scattering spectroscopy (ISS) data show Ti-rich termination. Fig. (b) and (c) are reproduced with permission from ref. **[80]**. Copyright 1994, American Association for the Advancement of Science.

While the recipe using the BHF based wet etching method for $SrTiO_3$ is well developed and in widespread use in the thin film community, it is pointed out that some oxygen atoms may be unintentionally replaced by fluorine atoms (as much as ~13%) at the surface during the etching process **[106]**. In this regard, it is noteworthy that other chemical treatment methods have been attempted as well for $SrTiO_3$ substrates; for example, Leca *et al.* annealed the perovskite substrates at 1000°C in $O_2$ atmosphere after etching the surface using different etching solutions (HCl+$HNO_3$ or HCl+$NH_4OH$) **[107]**. Irrespective of the kind of etching solutions, they always obtained atomically flat smooth substrate surfaces having $BO_x$ surface termination. Zhang *et al.* also used a HCl-$HNO_3$ (HCLNO) etching solution to produce atomically flat surfaces for $SrTiO_3$ (001) substrates ("Arkansas" method) **[108]**. Using depth-resolved cathodoluminescence spectroscopy (DRCLS), they proved that the extension of defect densities is higher for the BHF treated samples (hundreds of nm below the surface) compared to the HCLNO treated samples (less than 50 nm below the surface). Thus, the HCLNO treatment method seems to be more effective in producing an atomically flat surface with lower defect density. Although these etching and subsequent annealing methods have been successful to achieve atomically flat surfaces, it should be kept in mind that BHF or HCLNO etching solutions are highly acidic and thus toxic and hazardous from the safety point of view.

Alternative approaches without acidic solutions were also attempted to achieve atomically flat surfaces on $SrTiO_3$ substrates. Connell *et al.*, for example, followed a simple annealing and water leaching approach to produce an atomically flat surface. The treatment procedures include first thermal annealing (at 1000 °C for 1 h) of a substrate to induce SrO segregation and then dissolving Sr oxides into de-ionized water (DI-water leaching) **[81]**. Further annealing of the substrate at 1000 °C for 1 h in air produces an atomically flat surface. LFM measurements were used to confirm that the substrates are single terminated. Velasco-Davalos *et al.* adopted microwave-induced hydrothermal etching in DI-water for 3 min and subsequent thermal annealing at 1300 K for 10 min to produce an atomically flat surface. They also used LFM to confirm single termination **[109]**. Hatch *et al.*, on the other hand, attempted to compare the quality of the surfaces obtained by various wet etching methods, that is, BHF etching, HCl etching, and water leaching. Their XPS measurements revealed that the water leaching method is the most effective in removing $SrO_x$ crystallites and producing atomically flat surfaces with $TiO_2$-rich termination **[110,111]**. We may also mention that even annealing alone at high temperatures can produce an atomically-flat surface, but only with predominantly SrO termination **[112,113]** because strontium oxides segregate onto the surface from the bulk due to high temperature annealing **[114-116]**.

At this point it is noteworthy that the structure of the $SrTiO_3$ (001) surface is of scientific interest on its own, and there are several studies on the surface reconstruction using STM and other characterization methods. These surface science studies further reveal the topmost layer atomic arrangements and the possible termination of metal oxide substrates, and we cite briefly some of the reported cases. Matsumoto *et al.*, for instance, obtained the $(2 \times 2)$ surface reconstruction indicating O vacancy ordering in the $TiO_2$ topmost layer for the surface annealed in ultra-high vacuum (UHV) at 1200 °C **[117]**. Tanaka *et al.* observed the $(\sqrt{5} \times \sqrt{5})R26.6°$ surface reconstruction on $SrTiO_3$ (001) surfaces **[118]**; Kubo *et al.* also observed a similar surface reconstruction with $TiO_2$ termination **[119]**. Using LEED and high resolution STM, Jiang *et al.* showed the occurrence of the centered 6×2 $[c(6 \times 2)]$ and centered 4×2 $[c(4 \times 2)]$ reconstructions on $SrTiO_3$ (001) surfaces **[120]**, and Castell *et al.* also observed similar surface reconstructions for UHV annealed samples using high resolution STM and theoretical modelling **[121]**. Erdman *et al.* observed $(2 \times 1)$ and $[c(4 \times 2)]$ reconstructions on SrO deficient Ti-rich $SrTiO_3$ (001) surface **[122,123]**; Silly *et al.* also showed the formation of $TiO_2$ enriched $[c(4 \times 4)]$ reconstruction **[124]**. Using surface x-ray diffraction (SXRD), Herger *et al.* observed $(2 \times 1)$ and $(2 \times 2)$ surface reconstructions which can be best modelled with the $TiO_2$-rich surface layer **[125,126]**; Lin *et al.* also observed $(2 \times 2)$ surface reconstructions on $SrTiO_3$ (001) surfaces **[127]**. Recently, Dagdeviren *et al.* revealed the existence of $(\sqrt{13} \times \sqrt{13})R33.7°$, $[c(4 \times 2)]$, $(4 \times 4)$, $(2 \times 2)$, and $(2 \times 1)$ reconstructions on $SrTiO_3$ (001) surfaces by using LEED **[128]**. Their noncontact AFM and Auger electron spectroscopy (AES) measurements also showed Ti-enrichment on the surface. Interestingly with regard to the diverse surface reconstructions occurring on $SrTiO_3$ (001) surfaces, it was found that the relative stability of the different reconstructions can be altered by the addition of water on Ti-rich $SrTiO_3$ (001) surface **[129,130]**. Also there are several theoretical calculations that show that the surface with Ti-rich termination is the most energetically stable state **[131-133]**, and the Ti-rich termination was directly confirmed by high resolution transmission electron microscopy (HRTEM) **[134]**.

#### 3.2.1.2. $SrTiO_3$ (110) surface

$SrTiO_3$ (110) consists of alternate stacking of polar $SrTiO^{4+}$ and $O_2^{4-}$ layers with a step height of 2.76 Å as shown in **Fig. 5a**. For this orientation, the layers are polar unlike the (001) one and the polar nature of the layers strongly influences the surface state; either electronic reconstruction or stoichiometric modification (or atomic reconstruction) would occur on the surface to remove the surface polarization and stabilize the atomically flat structure as described above **[77-79]**. In addition, the fact that Sr and Ti cations appear in the same layer suggests that the relevant surface chemistry would be a little different from that of the (001) case. Mukunoki *et al.* reported the observation of an atomically flat $SrTiO_3$ (110) surface with one-unit cell step-terrace structure by annealing at a high temperature under varying oxygen partial pressures, and further showed that it is oxygen vacancies that stabilize the atomically flat surface [**135]**. Recently we demonstrated that atomically flat surfaces can be obtained for $SrTiO_3$ (110) via conventional BHF etching procedures as shown in **Fig. 5b** and **5c [104]**. We also showed, using TOF-MSRI measurements, that the topmost surface layer is indeed Sr-deficient and Ti-rich as presented in **Fig. 5d.** Theoretical calculations provided evidences that the compensation for the polar surface can be achieved through anomalous filling of surface states and keeping the insulating character of the surface with a stable Ti-rich layer **[136-138]**. Bachelet *et al.* observed that annealing substrates above 1000 °C produces atomically flat surfaces with a step height of two unit cells, called step bunching, for substrates with high miscut angle **[139]**.

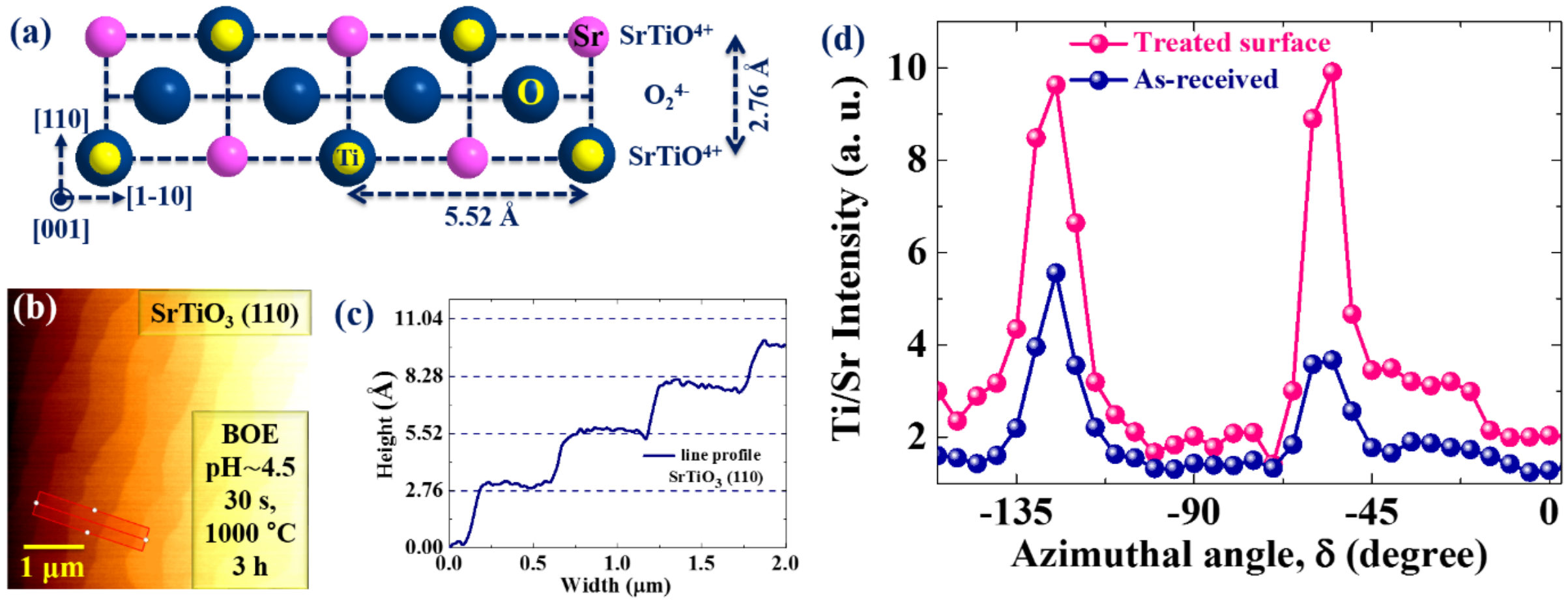

**FIG. 5:** (Color online) (a) Schematic side view of $SrTiO_3$ (110) showing alternating polar $SrTiO^{4+}$ and $O_2{}^{4-}$ layers with unit cell height of 2.76 Å. (b) Atomically flat surface showing one-unit cell step height obtained by the BHF method. (Inset indicates conditions of chemical etching and thermal annealing.) (c) Line profile along the line indicated in (b). The distance between the horizontal guide lines is the layer spacing expected for (110) orientation. (d) Time-of-flight mass spectroscopy shows that the surface is terminated by Sr-deficient and Ti-rich layer **[104].**

For $SrTiO_3$ (110), the exact chemical stability and surface reconstruction would be a more complex phenomenon, compared to the (001) case, due to the simultaneous presence of Sr and Ti cations on the topmost layer. Indeed, Bando *et al.* observed $(5 \times 2)$ and $c(2 \times 6)$ reconstructions on the surface **[140]**, while Wang *et al.* were able to obtain $(5 \times 1)$, $(4 \times 1)$, $(2 \times 8)$, and $(6 \times 8)$ surface reconstructions by tuning the Ti or Sr concentration **[141]**. On the other hand, Brunen *et al.* observed $(2 \times 5)$, $(3 \times 4)$, $(4 \times 4)$, $(4 \times 7)$, and $(6 \times 4)$ reconstructions by combining STM, LEED, and AES measurements **[142]**. By using similar techniques, Russell *et al.* showed the occurrence of $n \times 1$ $(3 \leq n \leq 6)$ surface reconstructions, where the surface is Ti-rich for $n$ = 3 and 4 and Sr-rich for $n$ = 6 **[143]**. Enterkin *et al.* also observed a variety of reconstructions $(n \times 1,\ n \geq 2)$ on the atomically flat $SrTiO_3$ (110) surface, using LEED and STM measurements, where the topmost surface layer is Ti-rich **[144]**.

### 3.2.1.3. $SrTiO_3$ (111) surface

$SrTiO_3$ (111) consists of alternate stacking of charged $SrO_3^{4-}$ and $Ti^{4+}$ layers with a unit cell step size 2.25 Å as shown in a schematic of **Fig. 6a**. The polar nature of the (111) orientation again offers difficulties in obtaining the atomically flat surface; however, Sr and Ti cations appear in separate layers and the surface chemistry would be similar to that of the (001) case. There are several studies available in the literature on producing the atomically flat surface for $SrTiO_3$ (111) by annealing only. Tanaka *et al.*, for example, reported that simple annealing at a high temperature produces a flat surface and subsequent analyses by STM measurements showed that annealing at 1180 °C forms a $SrO_{3-x}^{4-}$ topmost layer, whereas annealing at 1220 °C leads to a Ti topmost layer **[145]**. On the other hand, Sekiguchi *et al.* used CAICISS measurements on air annealed samples to suggest that the $Ti^{4+}$ layer is more stable than the $SrO_3^{4-}$ layer on the $SrTiO_3$ (111) surface **[146]**.

Turning to wet etching treatments of $SrTiO_3$ (111), Chang *et al.* successfully prepared an atomically flat surface with one-unit cell step-terrace structure using the conventional BHF etching method **[147]**. The single termination was confirmed by LFM analyses, and the AFM results are presented in **Fig. 6b** and **6c**. We also used the wet etching method to obtain atomically flat surfaces on $SrTiO_3$ (111) and confirmed the chemical nature of the surface termination using TOF-MSRI measurements. Our results shown in **Fig. 6d** reveals that the termination layer is Ti-rich **[104]**. In this case, however, the polar nature of the surface must be compensated for the Ti-rich layer to be stable. It is noted that Blok *et al.* also obtained the atomically flat surface in a similar way **[148]**; more importantly, however, they have shown that by growing a conductive buffer layer ($SrRuO_3$) one can maintain the bulk-like termination at the surface and grow other metal oxides coherently. Saghayezhian *et al.* demonstrated that there exists a critical annealing condition under which the Ti to Sr ratio at the surface becomes maximum **[149]**. Alternatively, a HCl-$HNO_3$ solution may be used as an etchant to produce the atomically flat surface, but with mixed termination **[150]**.

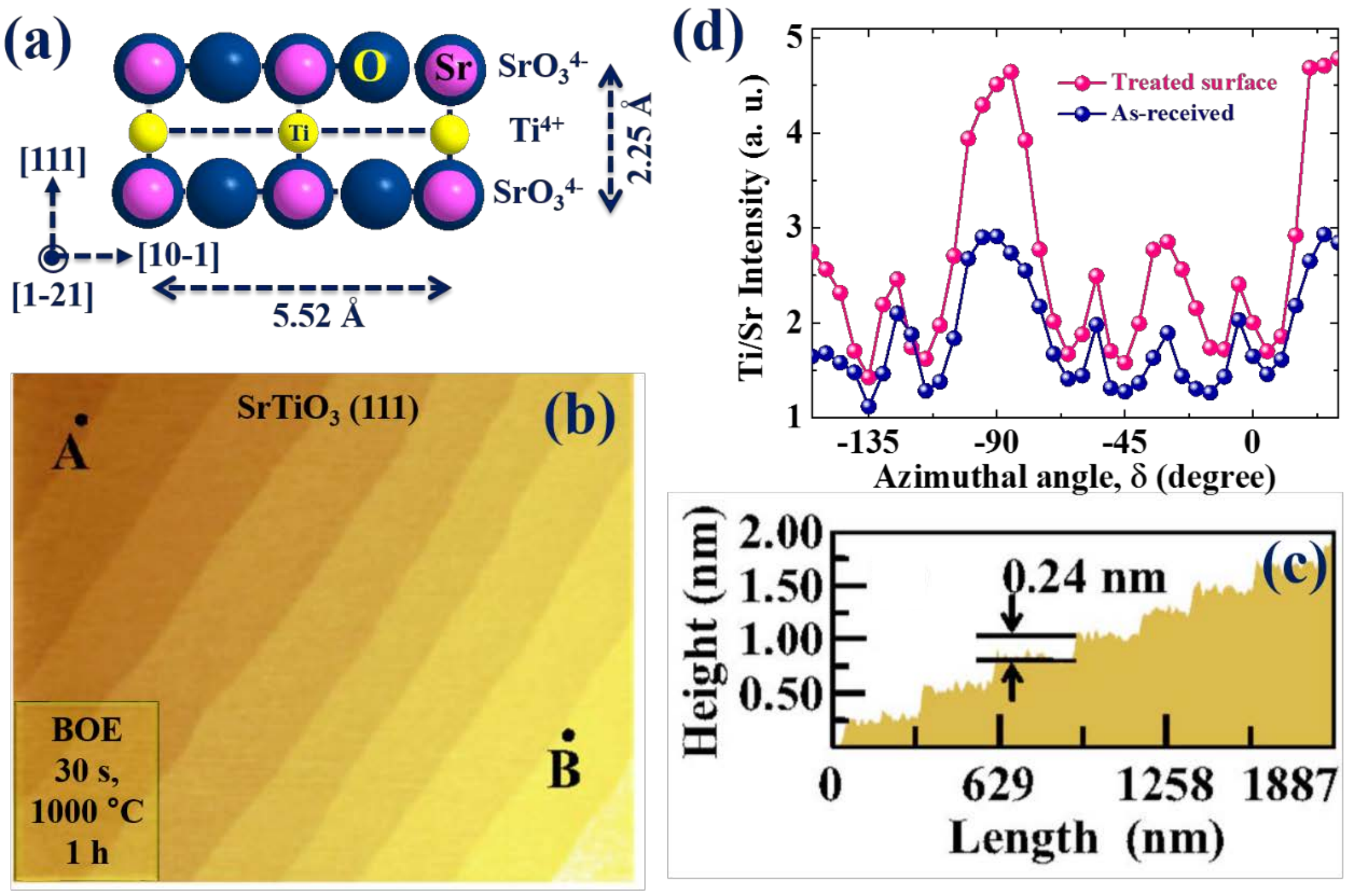

**Fig. 6.** (Color online) (a) Schematic side view of $SrTiO_3$ (111) surface showing that the unit cell consists of alternating polar layers of $SrO_3^{4-}$ and $Ti^{4+}$. (b) Surface morphology of a treated (111) surface obtained by atomic force microscopy. (c) Line profile showing unit cell step heights. (d) Atomically flat surface shows Ti-rich termination. Fig. (b) and (c) are adapted with permission from ref. **[147]**. Copyright 2008 AIP Publishing LLC.

For the (111) orientation, we naturally expect some kind of surface reconstructions to occur in order to compensate for the surface polarity. Indeed, Russell *et al.* observed via STM, LEED, and AES analysis that various surface reconstructions, i.e., $(\sqrt{7} \times \sqrt{7})R19.1°$ and $(\sqrt{13} \times \sqrt{13})R13.9°$, occur for samples annealed in UHV. They also showed that top few layers are enriched in Ti and Sr, but deficient in O, to compensate for the surface polarity **[151]**. They further showed that varying the oxygen partial pressure during annealing would bring about different $(n \times n)$ reconstructions with Ti-richness for the $SrTiO_3$ (111) surface **[152]**. Marks *et al.* also observed $(n \times n)(2 \leq n \leq 4)$ reconstructions on the Ti-rich (111) surface, using a combination of transmission electron diffraction, density functional theory modelling, and STM **[153,154]**. It is noteworthy that a theoretical modelling study already predicted that the stable low energy surface would be a Ti-rich layer for polar $SrTiO_3$ (111) **[136]**. Based on all the surface treatment methods and surface characterization studies, it would be safe to say that atomically flat $SrTiO_3$ surfaces generally show predominantly Ti-rich termination regardless of the orientations.

### 3.2.1.4. Role of the miscut angle

In addition to optimizing the treatment conditions for a substrate **[155],** the task of obtaining an atomically flat surface of a substrate requires consideration of another factor, the miscut angle of the substrate **[156]**. In the surface treatments of $SrTiO_3$ substrates, step bunching is frequently observed for high miscut ones. Here we wish to add a brief discussion on the miscut angle and step bunching despite the fact that it is general and not confined to the case of $SrTiO_3$. The miscut angle ($\theta$) of a substrate surface is defined as the angle between the surface and the crystal plane, which is equivalent to the angle between the surface normal ($S_n$) to the crystal plane normal ($C_n$) as drawn in **Fig. 7**. The miscut angle of the surface state, affected strongly the final surface characteristics after it was processed according to the etching and annealing method. For an ideally single-terminated surface, the miscut angle '$\theta$' and the average distance '$s$' between two adjacent steps are related as $\theta \sim l/s$, where '$l$' is the step height. The miscut angle of substrates strongly influences step heights and terrace widths; this is an intuitive result because as the miscut angle increases, the average terrace width must decrease if the step height is to be one-unit cell. Therefore, optimization of terrace width and step height should be conducted concurrently. Miscut angle of substrates would also affect the surface roughness and thin film growth morphology **[157-159]**.

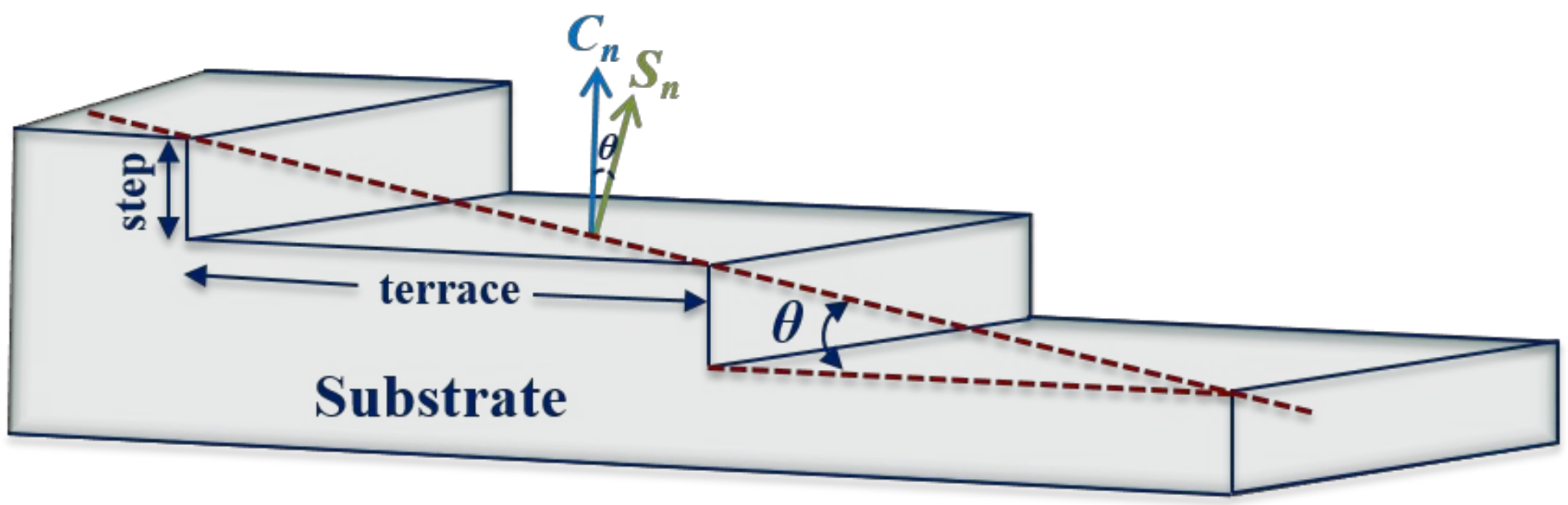

**Fig. 7.** (Color online) Schematic representation of a substrate. Miscut angle ($\theta$) of a substrate, defined as the angle between surface normal ($S_n$) and the crystal plane normal ($C_n$), determines step height and terrace width. It is seen that the step height would increase with an increase in miscut angle.

For substrates with high miscut angle, surface treatment procedures often bring about step bunching, which in turn results in steps with height larger than one-unit cell. The occurrence of step bunching for high miscut angle substrates may be understood simply in terms of minimization of surface free energy. According to theoretical considerations **[160-163]**, the surface free energy per unit area or surface tension $\gamma$ ($\theta$, $T$) can be written as a function of $\theta$ and temperature $T$:

$$\gamma(\theta,T) = \gamma^0(T) + \beta(T)\left(\frac{\tan\theta}{h}\right) + g(T)\left(\frac{\tan\theta}{h}\right)^3 \qquad (1)$$

in powers of the step density tan $\theta/h$ on the reference plane, where $h$ is the step height, and $\gamma^0(T)$ is the free energy density of the reference plane that represents the surface tension of the terraces between the steps. The second term represents the contributions from the steps, where $\beta(T)$ is the free energy per unit length for an isolated step. The third-order term arises from step-step interaction on the surface; this interaction consists of contributions from entropy repulsion as well as elastic and electrostatic interactions. These interaction contributions give rise to the third power of the step density; $g(T)$ is inversely proportional to the square of the distance between the steps. As $\theta$ increases, the interaction term becomes increasingly important in determining the surface structure. To minimize the surface energy of substrates with high $\theta$, the adjacent steps bunch by self-organization within the plane to increase the number of unit cells per step. This bunching effect becomes predominant as $\theta$ increases; as a result, the step height increases. Thus, it is always preferable to use low miscut angle ($\theta$ < 0.1°) substrates to obtain uniform step-terrace structures of one-unit cell step height.

### 3.2.2. *RE*ScO$_3$ (*RE* = Dy, Gd, and Nd) (110) surfaces

Rare earth (*RE*) scandates, *RE*ScO$_3$ with *RE* = Dy, Gd, and Nd, have an orthorhombic distorted structure that consists of alternately stacked polar *RE*O$^+$ and ScO$_2^-$ layers. The three *RE* scandates fall into a unique category of substrates because of their relatively large lattice constants; the pseudo-cubic lattice constant for the *RE* scandates ranges from 4.00 Å to 3.94 Å **[164]**. Due to the relatively large lattice constant, strain provided by *RE* scandate substrates allows emergent phenomena in thin films, which do not occur in bulk samples, of certain TMOs; for instance, ferroelectricity is seen in strained thin films of quantum paraelectric SrTiO$_3$ and ferroelectric ferromagnetism appears in strained EuTiO$_3$ thin films **[165,166]**.

Dirsyte *et al*. showed that atomically flat surfaces could be obtained for $DyScO_3$ (110) by only annealing at 1320 K in $O_2$ or Ar atmosphere. By using AES, they confirmed that the surface of $DyScO_3$ becomes DyO terminated after annealing in oxygen flow for 30–60 min, whereas annealing for a much longer time (~600 min) in an inert or oxygen atmosphere results in $ScO_x$ termination **[167]**. The reasoning for these peculiar behaviors offered by the authors is as follows: DyO termination in the oxygen annealing case would be due to the successive diffusion events, i.e., diffusion of oxygen from the ambient gas into the bulk, oxidation of dysprosium from $Dy^{3+}$ to $Dy^{4+}$, evaporation of Dy from the surface, subsequent diffusion of Dy from the bulk towards the near-surface layers, and consequent surplus of Dy on the surface and DyO termination. On the other hand, annealing for a long time in inert atmosphere leads eventually to the depletion of Dy from the surface and consequent $ScO_x$ termination because of the intrinsic bond strength difference, that is, the electrostatic bonding of Dy-O is weaker than the Sc-O bonding.

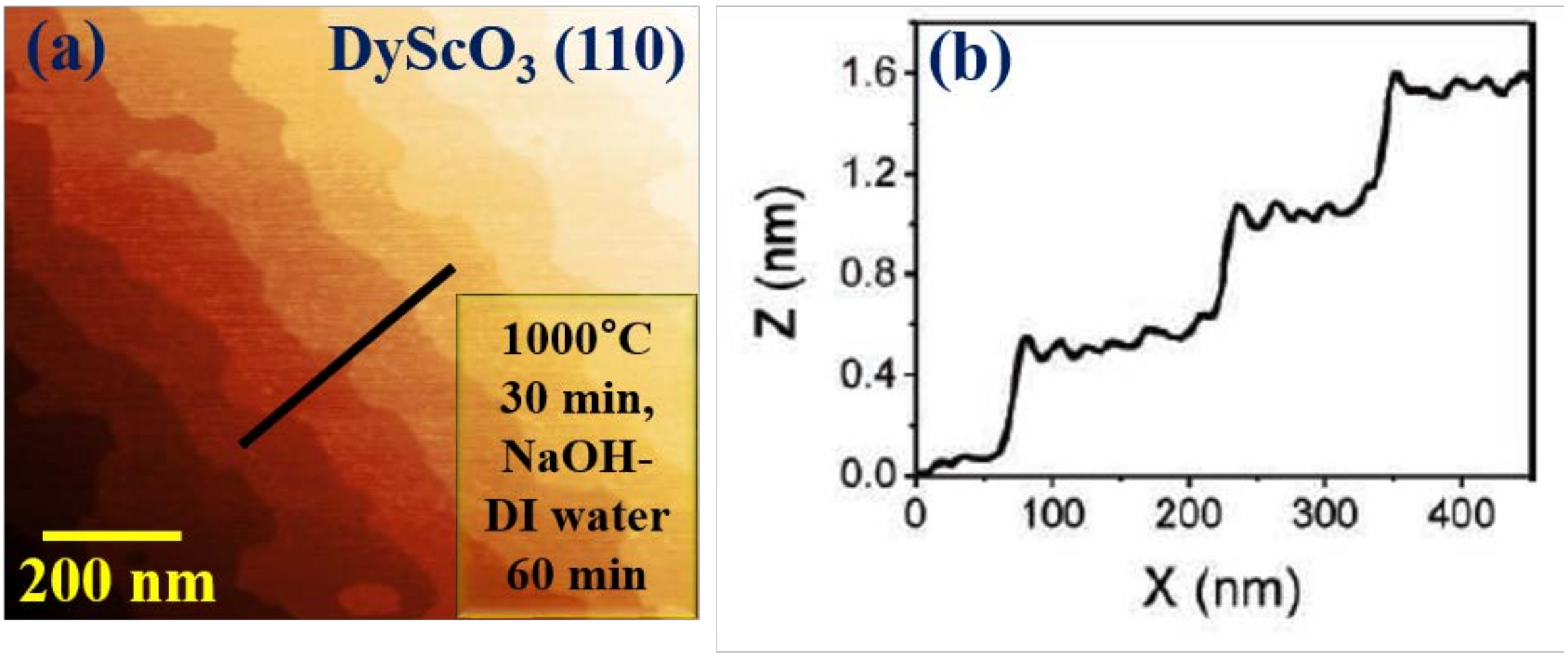

**Fig. 8.** (Color online) (a) AFM image of atomically flat scandate substrate ($DyScO_3$) obtained by annealing and subsequent basic solution wet etching treatments. Inset of (a) shows the annealing and wet etching treatment conditions. (b) Line profile along the black line in (a); step height is that of one-unit cell (~ 4 Å). Adapted with permission from ref. **[168]**. Copyright © 2010 WILEY-VCH Verlag GmbH & Co. KGaA, Weinheim.

Recently, Kleibeuker *et al*. showed that atomically flat surfaces could also be achieved on the three *RE* scandates by adopting a basic solution of NaOH, as opposed to using an acid etching solution, as a wet etchant **[168,169]**. As-received substrates are annealed at 1000 °C to induce segregation of one of the elements (e.g., Dy in the case of $DyScO_3$ because its bond strength is weaker than that of Sc) onto the surface. Then the surface layer of the annealed substrates is etched out by immersion in 12 M NaOH-DI water solutions. The one-unit cell step height of ~4 Å of such treated surface was confirmed by AFM measurements as shown in **Fig. 8**. The TOF-MSRI analyses revealed that the surfaces were fully $ScO_2$ terminated for all three scandates as shown in **Fig. 9 [168]**. It is noted that, due to the polar nature of the surface of the scandates, there might occur polar reconstructions, oxygen vacancies, or adsorbents on the surface to achieve a stable topmost layer. This stable B-site termination was further confirmed by the same group with the observation of $(1 \times 1)$ reconstruction, by combining RHEED and SXRD **[169]**. The most important fact is that this wet basic etching method is able to produce smooth atomically flat one-unit cell step-terrace surface with predominant B-site termination, $ScO_2$.

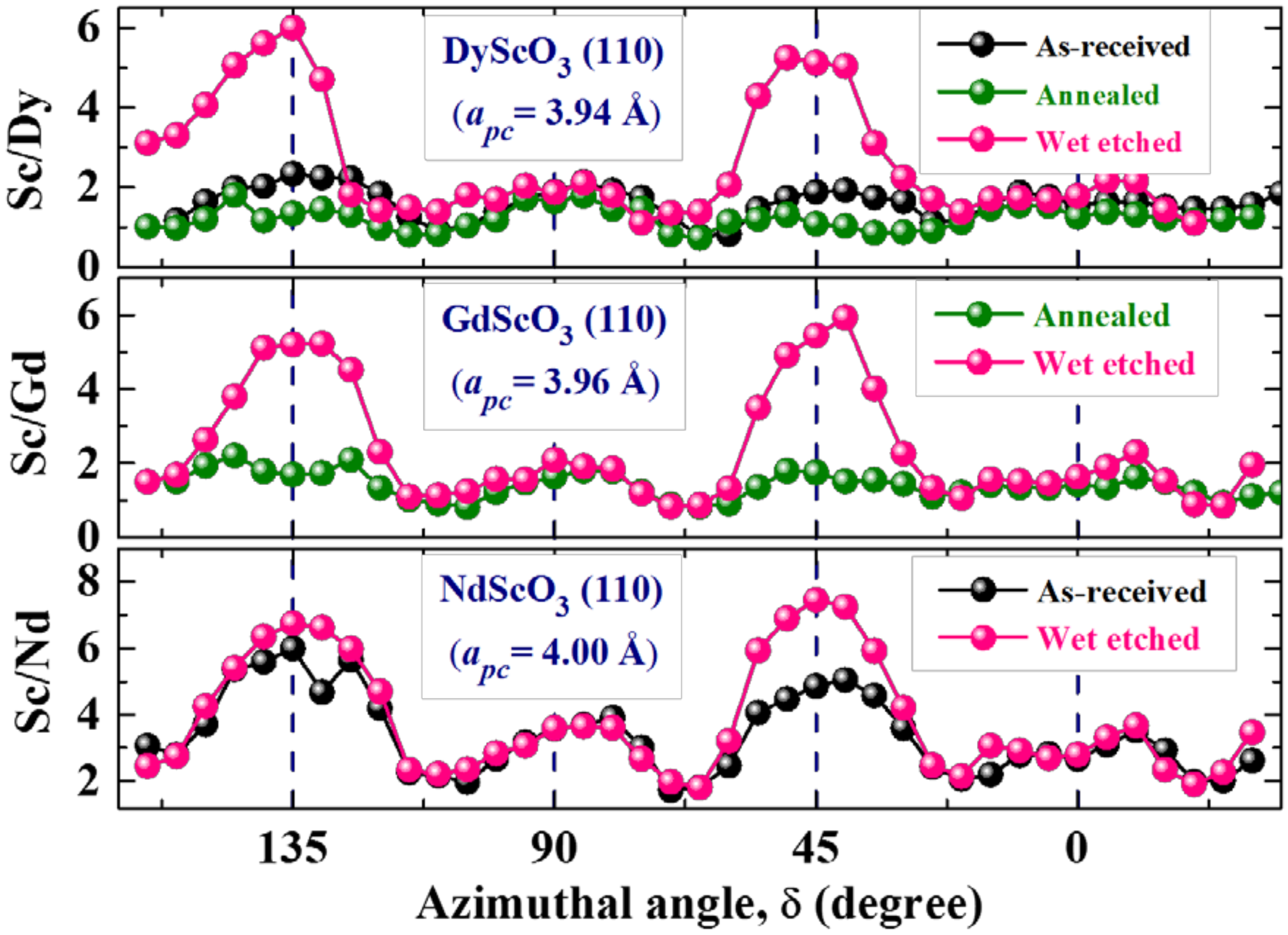

**Fig. 9.** (Color online) Intensity ratio of Sc versus rare earth (*RE*) element measured by TOF-MSRI spectrometry for $DyScO_3$, $GdScO_3$, and $NdScO_3$ substrates. Depending on the atomic structure, the wet-etched substrates show clear ratio maxima at 45° and 135° arising from $ScO_2$ termination. Adapted with permission from ref. **[168]**. Copyright © 2010 WILEY-VCH Verlag GmbH & Co. KGaA, Weinheim.

### 3.2.3. $NdGaO_3$ (001) surface

Neodymium gallate, $NdGaO_3$, forms a distorted orthorhombic structure with space group *Pbnm* at room temperature as shown in **Fig. 10a**; the orthorhombic lattice parameters are $a$ = 5.42817 Å, $b$ = 5.49768 Å, and $c$ = 7.70817 Å **[170,171]**. In general, an orthorhombic structure may be viewed conveniently as a cubic one: orthorhombic (*o*) [001] and [110] directions correspond to the pseudo-cubic (*pc*) [001] and [010] directions, respectively, as indicated in **Fig. 10a**. For $NdGaO_3$, the pseudo-cubic lattice constant is $a_{pc}$ ~ 3.86 Å. The pseudo-cubic structure consists of alternate stacking of polar layers, $GaO_2^-$ and $NdO^+$ as shown in **Fig. 10b**. The distance between two successive $NdO^+$ or $GaO_2^-$ layers is again $a_{pc}$ ~ 3.86 Å, corresponding to the one-unit cell step height.

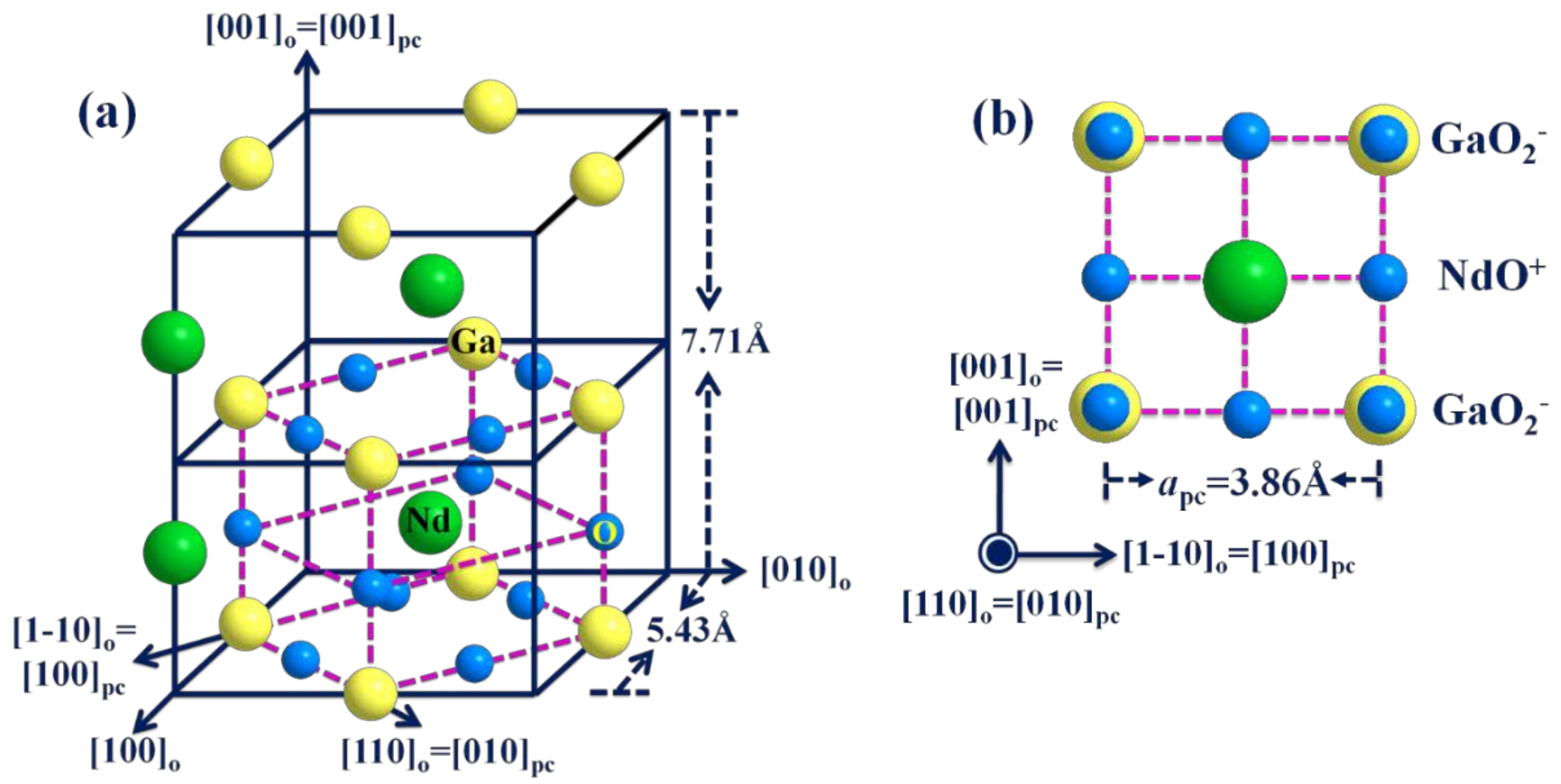

**Fig. 10.** (Color online) (a)-(b) Schematic representation of the orthorhombic perovskite unit cell of $NdGaO_3$. The pseudocubic unit cell (broken magenta lines) has $a_{pc}$ ~ 3.86 Å with alternating layers of polar $GaO_2^-$ and $NdO^+$.

Atomically flat A-site termination on $NdGaO_3$ (001) substrates was achieved by Ohnishi *et al.* via pure annealing treatment **[172]**. The AFM image of the treated $NdGaO_3$ (001) surface is shown in **Fig. 11**; the chemical status of the surface was confirmed to be Nd-rich (A-site) by CAISISS measurements. Gunnarsson *et al.* also obtained atomically flat surfaces by annealing only and showed Nd-rich termination using XPS measurements **[173]**. Radovic *et* *al.* demonstrated that a constant flow of pure oxygen at 1200 ° C for 20 h also produces atomically flat surfaces with single termination for $NdGaO_3$ (001) substrates, again confirmed by XPS **[174]**; Talik *et al.* also showed the A-site termination using a similar method **[175]**. AES measurements were also used to confirm the NdO termination for annealed surfaces, where Ga and O are depleted from **[176]**. Cavallaro *et al.* annealed the substrate at a high temperature of 1000 °C under different gas fluxes and showed that, irrespective of various gas types, $NdGaO_3$ surfaces always end up with the A-site rich termination as verified by Low Energy Ion Scattering (LEIS) **[177]**. Thermal annealing seems to allow the evaporation of gallium ions from the surface by forming volatile species $Ga_2O$. In order to alter the surface termination, solutions of BHF or HCl+$NH_4Cl$ were tested as an etchant and the result was analyzed by RHEED; indeed, B-site termination was successfully achieved by the etching method on $NdGaO_3$ surfaces **[107,178]**. From these various studies on $NdGaO_3$ (001) substrates, it is now established that high temperature annealing is sufficient to produce A-site terminated atomically flat smooth $NdGaO_3$ surfaces.

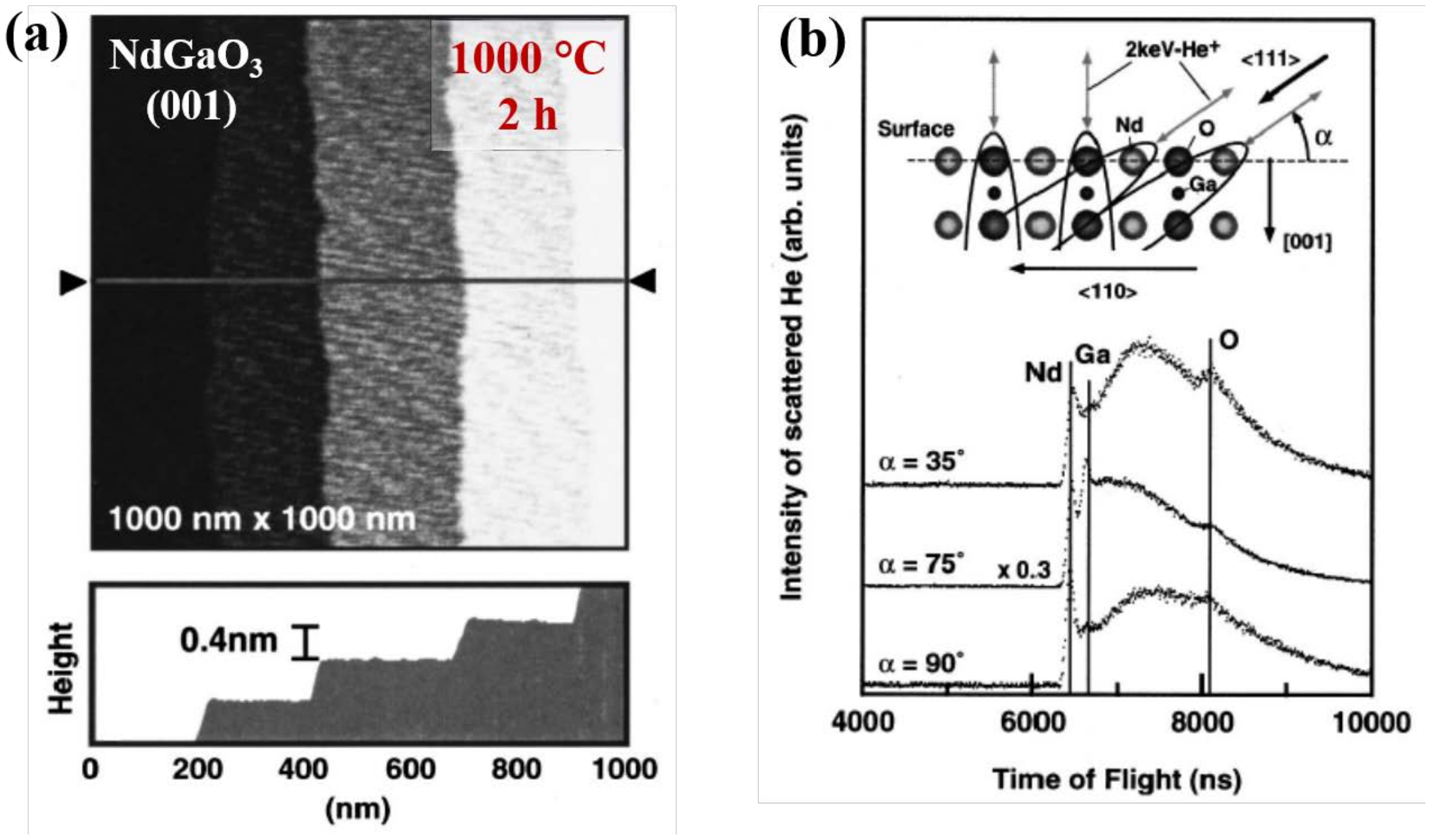

**Fig. 11.** (Color online) (a) (top) AFM surface image of $NdGaO_3$ substrate after annealing treatment. Thermal treatment was applied at 1000 °C in air for 2 h. (bottom) Step-terrace structures are seen with the distance between two horizontal lines ~4 Å. (b) Time of flight spectra show Nd-rich termination. Reproduced with permission from ref. **[172]**. Copyright 1999 AIP Publishing LLC.

### 3.2.4. $LaAlO_3$ (001) surface

$LaAlO_3$ has a rhombohedral structure ($a = b = c = 3.78$ Å, $\alpha = 90.066°$) with space group *R3mR* (No. 160) at room temperature **[68]**. It is a band insulator with a large band gap (~5.6 eV) and a reasonably high dielectric constant (~25). Single crystals of $LaAlO_3$ are readily available as substrates. The rhombohedral (or pseudo-cubic) crystal structure of $LaAlO_3$ consists of alternating polar layers of $LaO^+$ and $AlO_2^-$, and an ideal termination would consist purely of either $LaO^+$ or $AlO_2^-$. The distance between the adjacent $LaO^+$ and $AlO_2^-$ layers is ~3.78 Å, one-unit cell step height, as shown in **Fig. 12a**. It may be noted that $LaAlO_3$ became immensely popular as substrates for growing epitaxial $BiFeO_3$ thin films, which exhibit exotic multifunctional properties **[39]**.

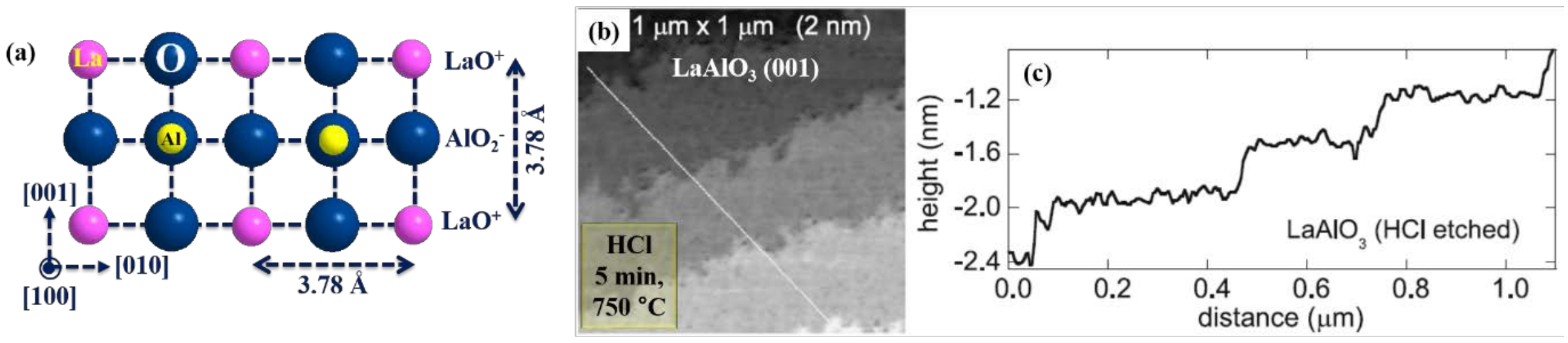

**Fig. 12.** (Color online) (a) Schematic side view of $LaAlO_3$ (001) unit cell showing the alternating stacking of $LaO^+$ and $AlO_2^-$ layers. (b) AFM image of a treated $LaAlO_3$ (001) surface. (c) Line scan result of one-unit cell step height ~4.0 Å and terrace width ~400 nm. Figure (b) and (c) were reproduced with permission from ref. **[173]**. Copyright © 2016 Elsevier B.V.

The most appropriate method to treat a $LaAlO_3$ substrate is found to be as follows: (1) Clean a substrate with acetone and methanol, and then soak in ultrasonically-stimulated water. (2) Wet etch for 30 s with dilute HCl (pH ~ 4.5). This pH of HCl is chosen to selectively etch out the lanthanum-related oxides to produce a surface that consists of single terminated $AlO_2$. (3) Anneal at 1000 °C for 2.5 h. The annealing produces an atomically flat surface with step-terrace width ~350 nm and one-unit cell step height ~3.78 Å as shown in **Fig. 12b-c [172,173,179]**. To identify the topmost atomic layer of treated $LaAlO_3$ (001) substrates, TOF-MSRI measurements were performed; the result shown in **Fig. 13** indicates that the topmost layer is of Al-rich termination, predominantly of $AlO_2$ **[172]**. Gunnarsson *et al.* also confirmed, by LFM and XPS analyses, that HCl treated $LaAlO_3$ (001) surfaces show predominantly $AlO_x$ termination **[173]**. For a $LaAlO_3$ single crystal annealed at 1200 °C in flowing oxygen, X-ray truncation rod analysis showed B-site termination of the surface with minor structural rearrangement such as relaxation of O and La atoms relative to the ideally terminated surface **[180]**. Kim *et al.* also reported that annealing a $LaAlO_3$ substrate at high temperatures (1000-1100 °C) for 10 hrs under the oxygen flow produces $AlO_2$ terminated surfaces, possibly due to the segregation of $AlO_2$ at the surface **[181]**.

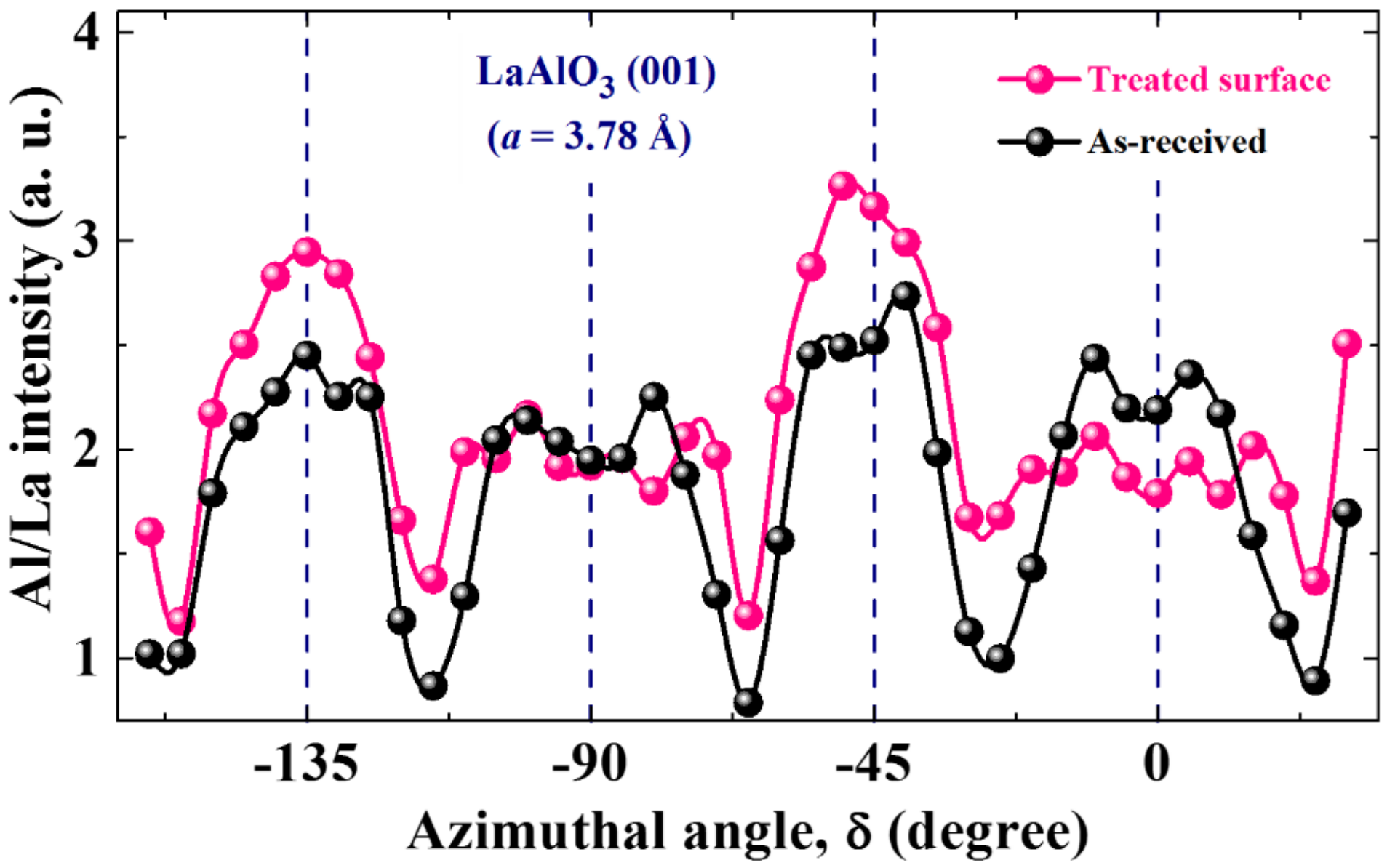

**FIG. 13:** (Color online) (a) TOF-MSRI counts of Al and La ions and intensity ratio profile of as-received and treated $LaAlO_3$ substrate vs. azimuthal angle. Al/La intensity ratio is highest at -45° and -135°; Al rich termination is predominant at the surface.

It should be pointed out that there are some mixed reports on the surface termination of $LaAlO_3$ in literature. In fact, the surface termination itself may depend on temperature; it was reported in literature that the topmost layer of $LaAlO_3$ varies from Al-O termination between room temperature to 150 °C to La-O termination above 250 °C while mixed termination is observed in the intermediate region of 150 ~ 250 °C **[182-184]**. Lanier et al. observed $(\sqrt{5} \times \sqrt{5})$ R26.6° surface reconstruction on the $LaAlO_3$ (001) surface, forming LaO termination with one lanthanum vacancy per surface unit cell to compensate for the charged nature of the surface **[185]**. Another study on $LaAlO_3$ (001) substrates annealed at 1100 °C, using low-energy electron microcopy, revealed a mixed termination layer on the surface, where LaO terminated regions are well reconstructed $(\sqrt{5} \times \sqrt{5})$R26 and $AlO_2$ terminated regions are unreconstructed **[186]**. Theoretical calculations suggest that depending on the surface reconstruction, $LaAlO_3$ surfaces can show either LaO or $AlO_2$ termination **[187]**. In short, there are still some controversies about the chemical termination and the surface reconstructions of $LaAlO_3$; nevertheless, it seems to be established that by adopting HCl based wet chemical etching, atomically flat surfaces can be obtained for $LaAlO_3$ (001) substrates with predominantly $AlO_2$ termination.

### 3.2.5. $YAlO_3$ (110) surface

Yttrium orthoaluminate $YAlO_3$ is a distorted perovskite composed of orthorhombic unit cells with space group *Pnma* at room temperature. $YAlO_3$ is a wide band gap (~7.9 eV) insulator with orthorhombic lattice parameters $a$ = 5.330 Å, $b$ = 5.180 Å, and $c$ = 7.375 Å. From these lattice parameters, the pseudo-cubic lattice constant $a_{pc}$ is obtained to be ~ 3.72 Å **[188,189]**. Liu *et al.* achieved atomically flat single terminated surfaces of $YAlO_3$ (110) by mere annealing at 1000 °C for 2 h. This simple treatment produced atomically flat step-terrace structures as shown in **Fig. 14** [**190**]. Caution must be exercised, however, because it was reported that annealing a $YAlO_3$ crystal at temperatures above 1160 °C would generate a second phase $Y_3Al_5O_{12}$, and annealing at an even higher temperature 1300 °C would cause segregation to $Al_2O_3$ and $YAlO_3$ **[191]**. The nature of the topmost terminating layer has not been identified yet, and further characterization is required.

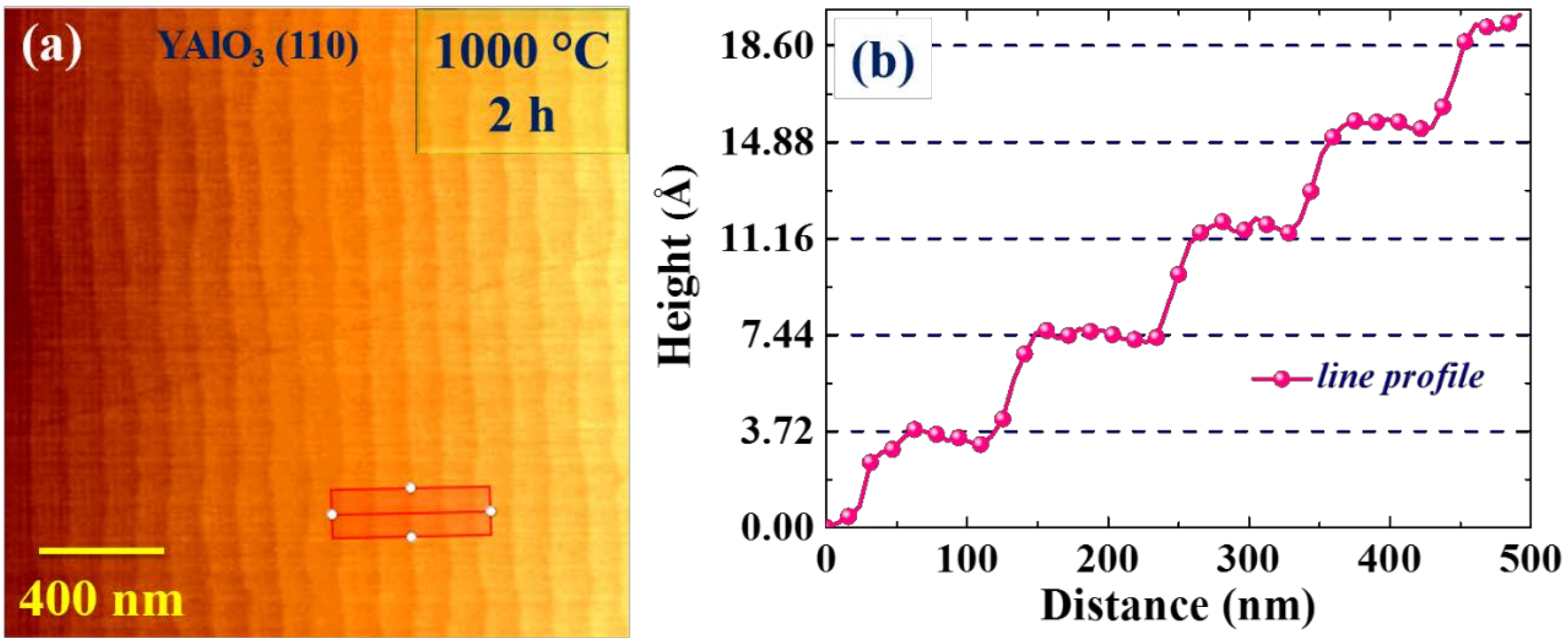

**Fig. 14.** (Color online) (a) AFM image of the surface of a treated $YAlO_3$ (110) substrate. Straight step-terrace structures are seen. Treatment procedure was mere annealing at 1000 °C

for 2 h. (b) Line profile of the red rectangle of figure (a): the step height corresponds to one-unit cell ~3.72 Å. Treatment procedure was taken from the ref. **[190]**.

### 3.2.6. $KTaO_3$ (001) surface

$KTaO_3$ is a prototypical $A^{1+}B^{5+}O_3$ cubic perovskite substrate with lattice constant ~3.9885 Å. Its dielectric constant is extremely high with ~4500. $KTaO_3$ substrates are useful for growing high quality oxide thin films [**192,193**] and, more importantly, for generating two dimensional electron gas with high mobility at room temperature **[194-196]**. Recently it was shown that superconductivity could be induced in $KTaO_3$ by electrostatic carrier doping **[197]**. Atomically flat surface can be obtained for $KaTaO_3$ (001), by wet etching with a commercially available BHF followed by annealing treatments in air **[198]**. Etching the surface for 15 min produces a step-terrace structure with step height of one-unit cell (~4 Å) and terrace width of 500-800 nm. It is presumed that hydroxides are removed by the acidic etchant, producing the single terminated atomically flat surface. Judging from the formability and solubility of KOH in a BHF solution, one can conjecture that the surface is predominantly terminated by $TaO_2$. However, there is no direct experimental evidence available yet for the chemical nature of the surface. It is noteworthy that Nakamura and Kimura noticed an interesting effect during the surface treatment of $KTaO_3$; that is, the field effect characteristics of $KTaO_3$, which is useful for field effect device applications and can be controlled by varying the HF concentration **[199]**.

## 3.3. Cation-rich thermal annealing treatments

For more complex metal oxide substrates with multiple cations such as $La_{0.18}Sr_{0.82}Al_{0.59}Ta_{0.41}O_3$ (LSAT) and $SrLaAlO_4$, thermal annealing under cation-rich environments, rather than straight thermal annealing, is found to be a unique way for obtaining atomically flat single terminated surfaces. In this section, we review the procedures for preparing the atomically flat surfaces for these substrates.

### 3.3.1. $La_{0.18}Sr_{0.82}Al_{0.59}Ta_{0.41}O_3$ (001) surface

$La_{0.18}Sr_{0.82}Al_{0.59}Ta_{0.41}O_3$ (LSAT) has a cubic structure with lattice constant of ~3.88 Å. LSAT substrate has become very popular for growing high $T_C$ superconducting films due to its well-matched lattice constant **[200]**. Chemically, however, it is a mixed perovskite which contains two kinds of cations each in A-sites as well as in B-sites **[201]**. This chemical complexity then is expected to make the system difficult to etch out and induce the system either A-site or B-site termination by conventional wet etching and/or thermal annealing treatments. Ohnishi *et al*. annealed the surface at 1300 °C in air and showed the dominant B-site termination **[172]**. Leca *et al*. adopted wet etching with a $HCl+HNO_3$ etching solution and annealed the substrate at 1000 °C for 3h in $O_2$ environment to produce an atomically flat surface **[107]**.

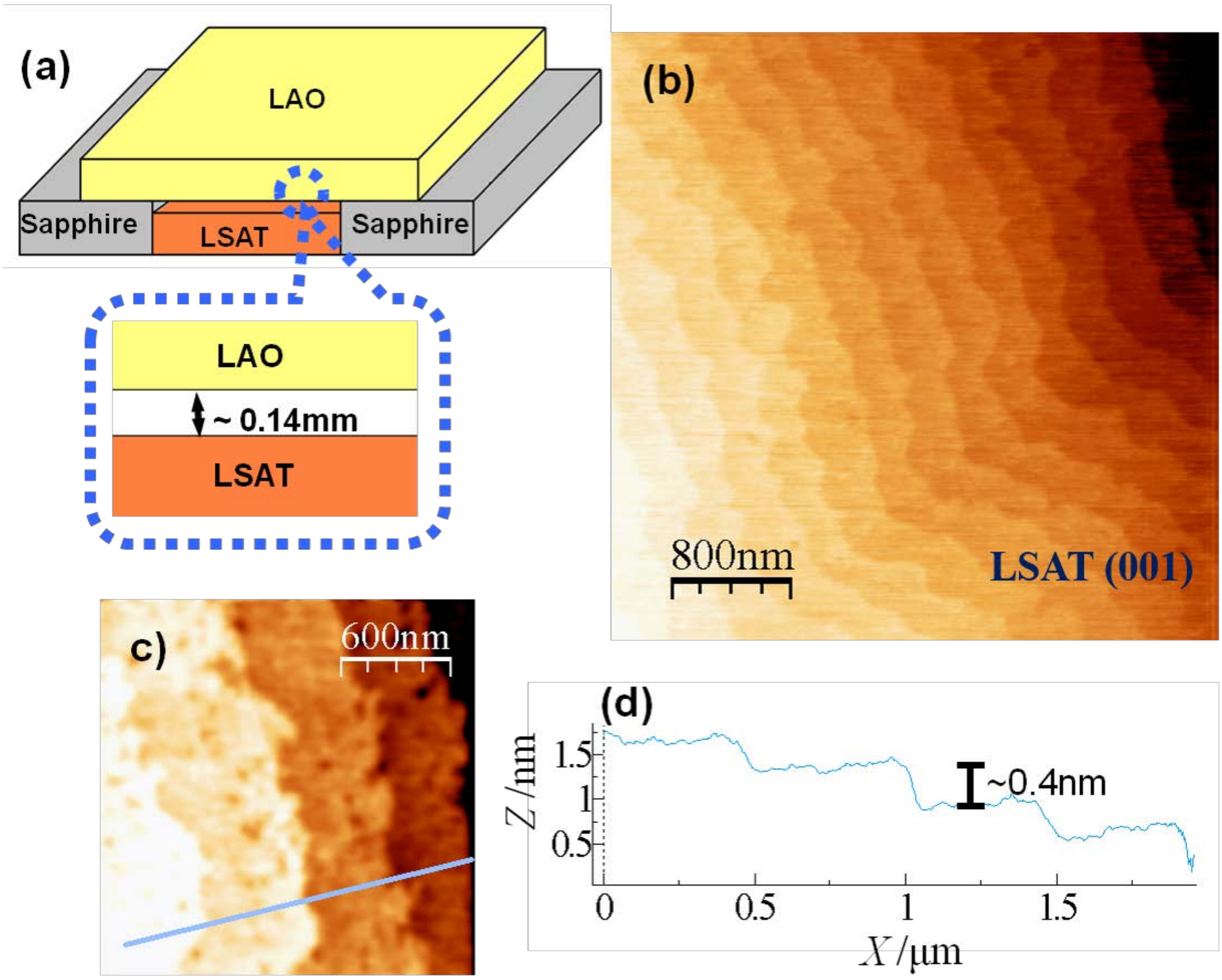

**Fig. 15.** (Color online) (a) Schematics of annealing environments to obtain atomically flat LSAT surface. $LaAlO_3$ (LAO) single crystal cover was used to provide additional La ions which were lost during annealing. (b)-(d) AFM images and line profile of atomically flat surface with one-unit cell step height ~4 Å. Reproduced with permission from ref. **[82]**. Copyright © 2010 WILEY-VCH Verlag GmbH & Co. KGaA, Weinheim.

Recently, a rather different approach, taking the chemical complexity of LSAT into account, has been attempted to produce atomically flat surfaces. An atomically flat surface for a LSAT substrate was achieved by annealing it under La vapor pressure; La vapor was provided by placing a $LaAlO_3$ (001) substrate close to the LSAT substrate during high temperature annealing at 1300 °C. The experimental setup and the annealing results are shown in **Fig. 15 [82]**. High temperature annealing of a LSAT substrate induces the formation of many SrO particles on the substrate surface. This result indicates that La ions would evaporate from the surface while Sr ions would not, considering the fact that the A-sites contain both La and Sr ions. One way to compensate for the loss of La ions is to provide extra La ions to the surface, and this can be done by exposing the substrate to La vapor released from an annealed $LaAlO_3$ crystal as illustrated in **Fig. 15a**. AFM images prove the one-unit cell step-terrace structure as shown in **Fig. 15b-d**, and ion scattering spectroscopy measurements also confirm that the top surface is predominantly A-site terminated. By contrast, LSAT substrates treated at temperatures less than 1200 °C possess poorly defined steps. It is noted that atomically flat surface was also obtained by only thermal annealing LSAT at a high temperature, but the chemical termination was reported to be of mixed nature **[202-204]**.

### 3.3.2. $SrLaAlO_4$ (001) surface

$SrLaAlO_4$ (SLAO) belongs to a somewhat different class than the metal oxide substrates discussed so far. These substrates all have the perovskite $ABO_3$ structure with an alternating stacking sequence of AO and $BO_2$ in the [001] direction; however, SLAO possesses the layered structure of $K_2NiF_4$ with repeating layers of --$AlO_2$-(Sr,La)O-(Sr,La)O-$AlO_2$-(Sr,La)O-(Sr,La)O-- along the [001] direction as drawn schematically in **Fig. 16**. (Sr,La)O planes may be regarded as the A-site layer and the $AlO_2$ planes as the B-site layer; the *c*-axis lattice constant of this tetragonal structure is 12.6362 Å and in-plane lattice constants $a = b =$ 3.7569 Å **[205]**. SLAO substrates have been frequently used in growing layered TMOs and, in particular, so-called 214 high $T_c$ superconductors **[206,207]**. Thus, the availability of atomically flat SLAO substrates would provide researchers new opportunities to grow high quality thin films of various functional oxides, especially layered materials.

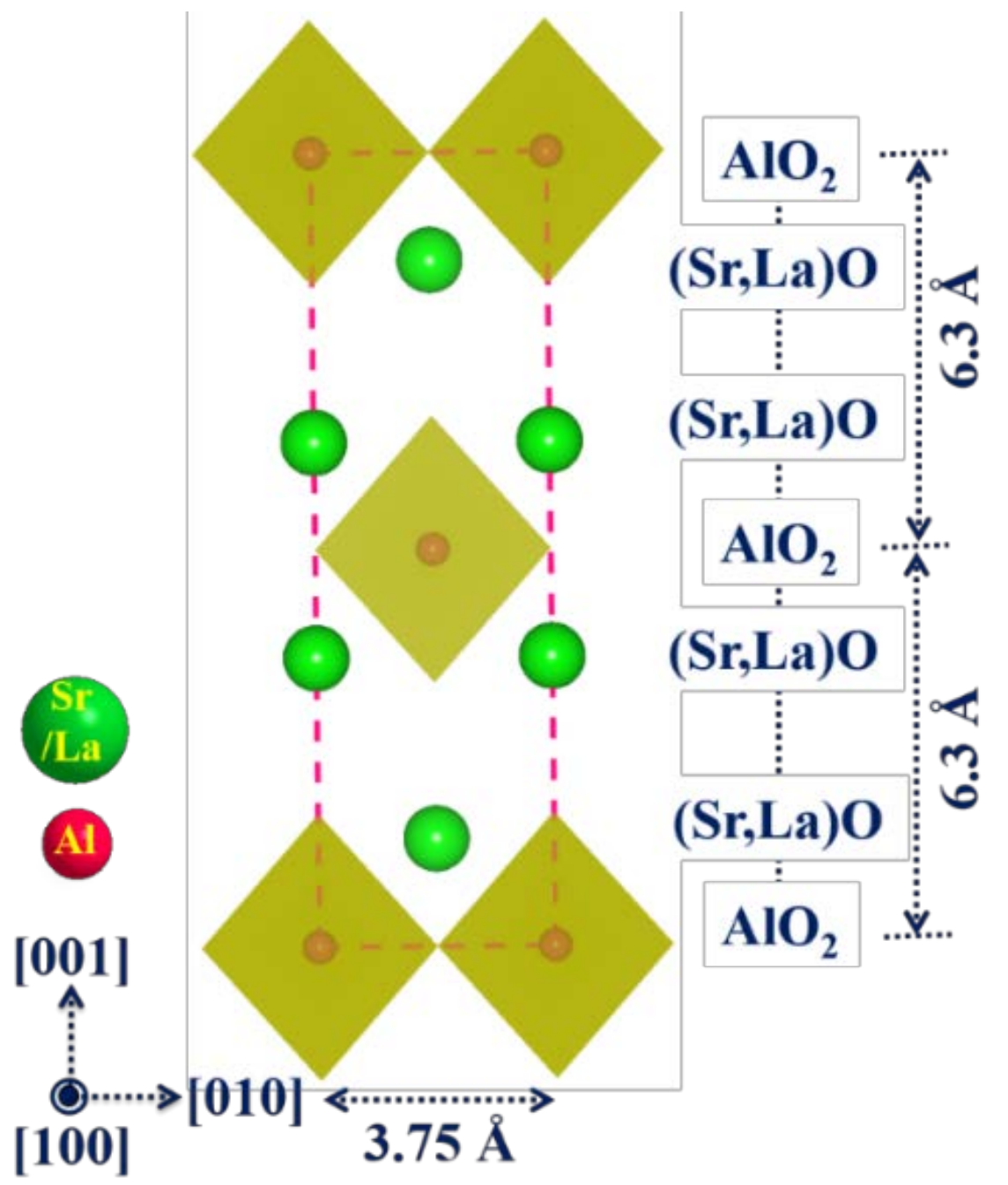

**Fig. 16.** (Color online) (a) Tetragonal $K_2NiF_4$ type crystal structure of $SrLaAlO_4$ with $a = b =$ 3.75 Å and $c$ = 12.6 Å. Along [001], the stacking sequence is -$AlO_2$-(Sr,La)O-(Sr,La)O-$AlO_2$-(Sr,La)O-(Sr,La)O- (the pink dotted rectangle denotes the unit cell). Crystal structure was drawn using VESTA software.

Preparing atomically flat surfaces of SLAO is not straightforward because of its unusual layered structure and also because both Sr and La cations are present in A sites; these characteristics would hinder the use of naive etching and annealing methods. Furthermore, the surface of SLAO is unstable against La loss at high temperatures, and annealing would cause the formation of SrO particles at the surface **[208]**. To avoid SrO formation, SLAO substrates should be annealed in cation rich environments such as the ones schematically depicted in **Fig. 17**. Indeed, atomically flat surfaces were successfully obtained with both setups; however, it should be noted that the chemical nature of the top surface varies depending on the annealing environment. Annealing a substrate in $La_2O_3$-rich environments yields a (Sr,La)O-terminated (A-site) surface whereas annealing in $LaAlO_3$ yields an $AlO_2$-terminated (B-site) surface. TOF-MSRI measurements confirmed that the layer sequence for $La_2O_3$ and $LaAlO_3$ treated SLAO substrates are (Sr,La)O-(Sr,La)O-$AlO_2$ and $AlO_2$-(Sr,La)O-(Sr,La)O from the top surface, respectively, as shown in Fig. **18 [209]**. The selective A-site or B-site termination obtained by this approach provides a useful pathway to obtain atomically flat surface without growing additional monolayers to alter the surface termination.

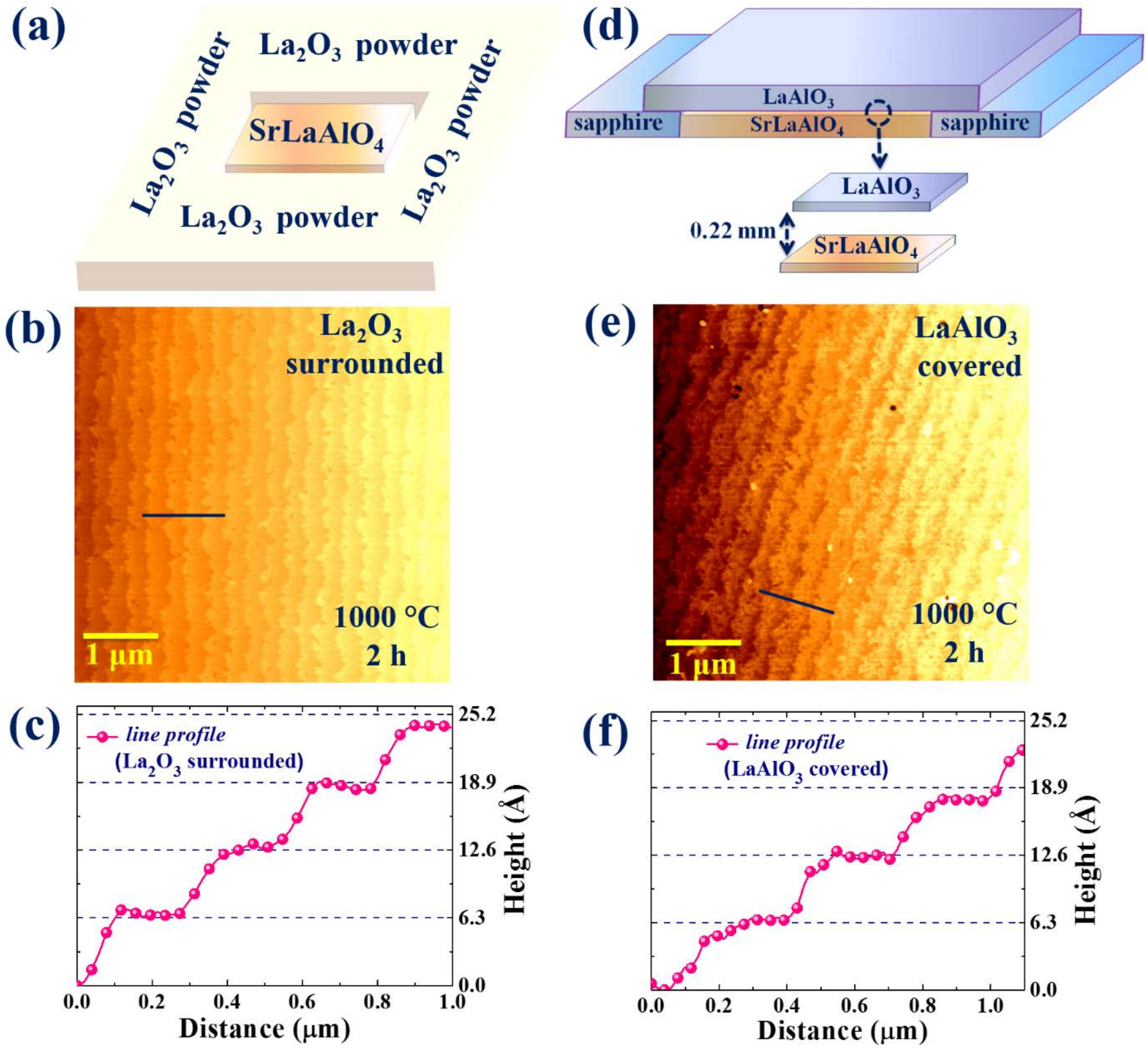

**Fig. 17.** (Color online) (a) Annealing setup for $SrLaAlO_4$ substrates surrounded by $La_2O_3$ powder. (b) AFM image of the surface after annealing at 1000 °C for 2 h. (c) Line profile along the dark line in (b). (d) Annealing setup for $SrLaAlO_4$ substrates with $LaAlO_3$ single crystal. (e) AFM image of the surface after annealing at 1000 °C for 2 h. (f) Line profile along the dark line indicated in (e). Step height corresponds to half unit cell spacing of 6.3 Å. Reprinted with permission from ref. **[209]**. Copyright 2013 AIP Publishing LLC.

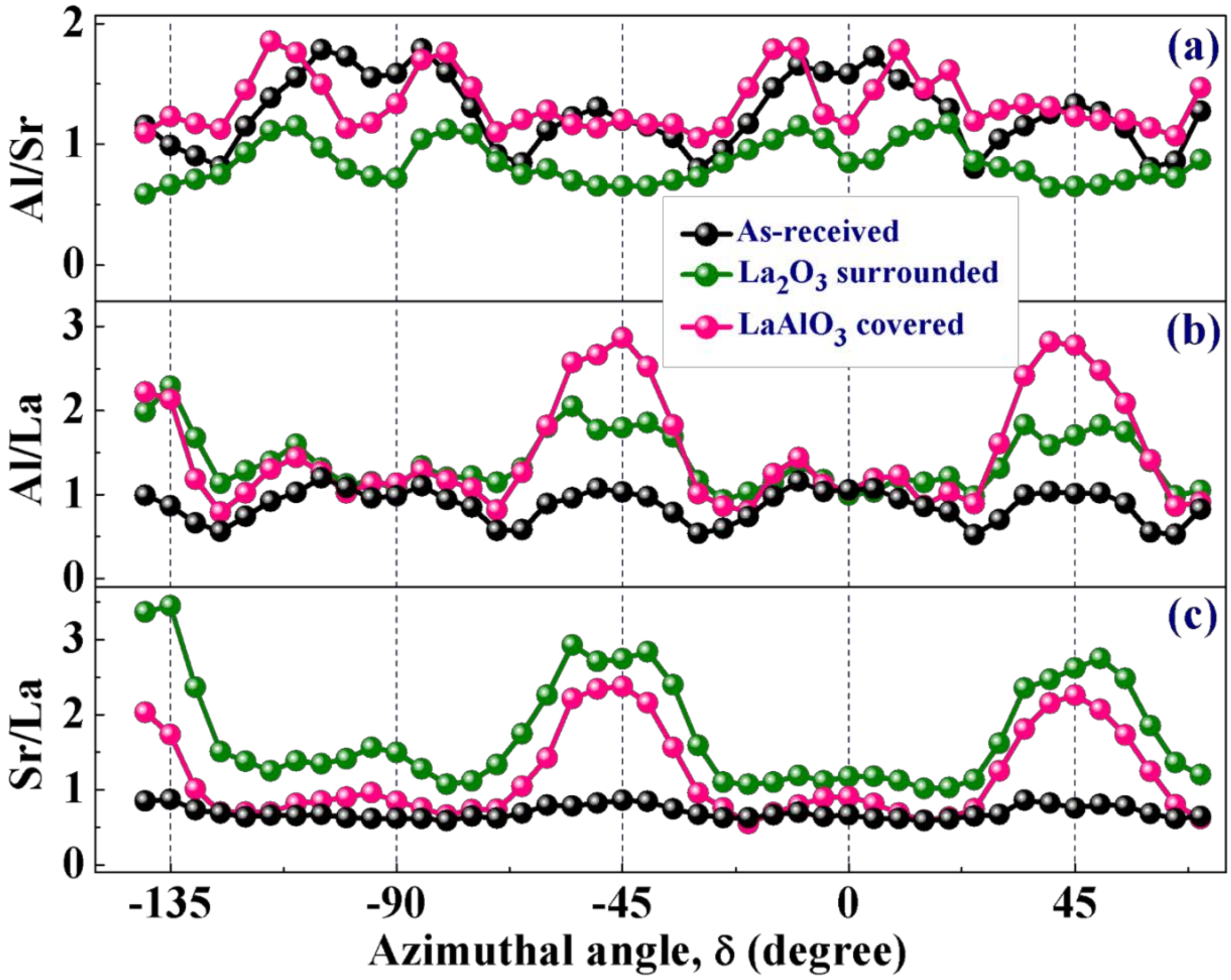

**Fig. 18.** (Color online) TOF-MSRI intensity ratios between (a) Al and Sr, (b) Al and La, and (c) Sr and La show the cationic arrangements of the top layers, and also reveal La-poor layers for annealed substrates. Reproduced with permission from ref. **[209]**. Copyright 2013 AIP Publishing LLC.

## 4. Conclusion and future prospects

Atomically flat single terminated substrates are a prerequisite for high quality atomically controlled layer-by-layer growth of epitaxial thin films and heterostructures. In the present review, we have covered the procedures for obtaining atomically flat and chemically homogenous surfaces of perovskite substrates with lattice constants ranging from 4.00 Å to 3.70 Å; various surface treatment methods typically adopt some combination of high temperature annealing and chemical etching. The treatment methods reviewed here include only thermal annealing at high temperatures, a combination of selective wet etching of complex oxides by acidic solutions and subsequent high-temperature thermal annealing, a combination of high-temperature thermal annealing and subsequent selective wet etching of complex oxides using basic solutions, high temperature thermal annealing in cation rich environments, etc. A fundamental understanding of the chemical processes associated with each substrate treatment to remove a particular type of metal oxides from the surface is nontrivial and perhaps far from being completely understood at present; thus, we have focused on the empirical and particularly practical aspects of the surface treatment procedures. Representative examples of the treated surfaces are collectively shown in **Fig. 19** as a summary. Ion scattering measurements revealed that depending on surface chemistry, the topmost layer of the treated substrates is predominantly either A- or B-site terminated; the termination characteristics of treated atomically flat substrate surfaces are summarized in Table **2** for quick reference. Perhaps a few words of caution are in order here. Although the values for treatment conditions indicated in the insets of **Fig. 19** are taken from the literature, there remains a possibility of variation in actual values. Exact temperatures, for example, may vary due to different sensors in annealing setups or even due to the variation in the position of the sensor because diffusion processes are exponentially dependent on temperature. Furthermore, high temperature annealing, particularly annealing for a long time in a tube furnace would produce point defects such as vacant lattice sites and interstitial atoms or ions, foreign atoms or ions in either interstitial or substitutional positions, dislocations, and most importantly oxygen vacancies on the substrate surface, which has strong influence on the exotic interface physics of thin film heterostructures **[210-212]**.

Although we have collected the surface treatment procedures for most commercially available perovskite-type metal oxide substrates, there still remain many areas to be explored in the future. First of all, expanding the list of substrates beyond the specified lattice constant range is desired for the progress of thin film growth as imposed strain due to lattice mismatch would pose a possibility of new functional properties not observed in their natural states. Smaller lattice parameter substrates would include ones such as $CaYAlO_4$ ($a$ = 3.645 Å), $LuAlO_3$ ($a$ = 3.675 Å), and $CaNdAlO_4$ ($a$ = 3.684 Å) while some larger lattice parameter examples are $LaScO_3$ ($a$ = 4.050 Å), $LaLuO_3$ ($a$ = 4.183 Å), MgO ($a$ = 4.212 Å), etc **[68,213,214]**. Atomically flat single terminated surfaces of these substrates would be immensely helpful to thin film researchers because each of them has its own advantage for epitaxial thin film growth. Binary oxide MgO, for example, has a simple cubic structure and has a very large lattice constant, much larger than that of the substrates covered in this review. At present it appears that the treatment method for obtaining atomically flat MgO surface is not firmly established **[215]**. In contrast to binary oxides, mixed perovskite materials (AA')(BB')O3 with multiple metal cations, such as $Nd_{0.4}Sr_{0.6}Al_{0.7}Ta_{0.3}O_3$ ($a$ = 3.835 Å) and $La_{0.15}Sr_{0.85}Ga_{0.575}Nb_{0.425}O_3$ ($a$ = 3.932 Å), may also be used as substrates **[216]**; they would deepen the pool of substrates and make a selection appropriate for a given situation a lot easier. These crystals, however, have more than one type of elements in both A- and B-sites (similar to LSAT), and the atomic composition and arrangement in the materials along the [001] direction as well as other directions are more complex. It would be rather challenging to obtain atomic flatness on these substrate surfaces.

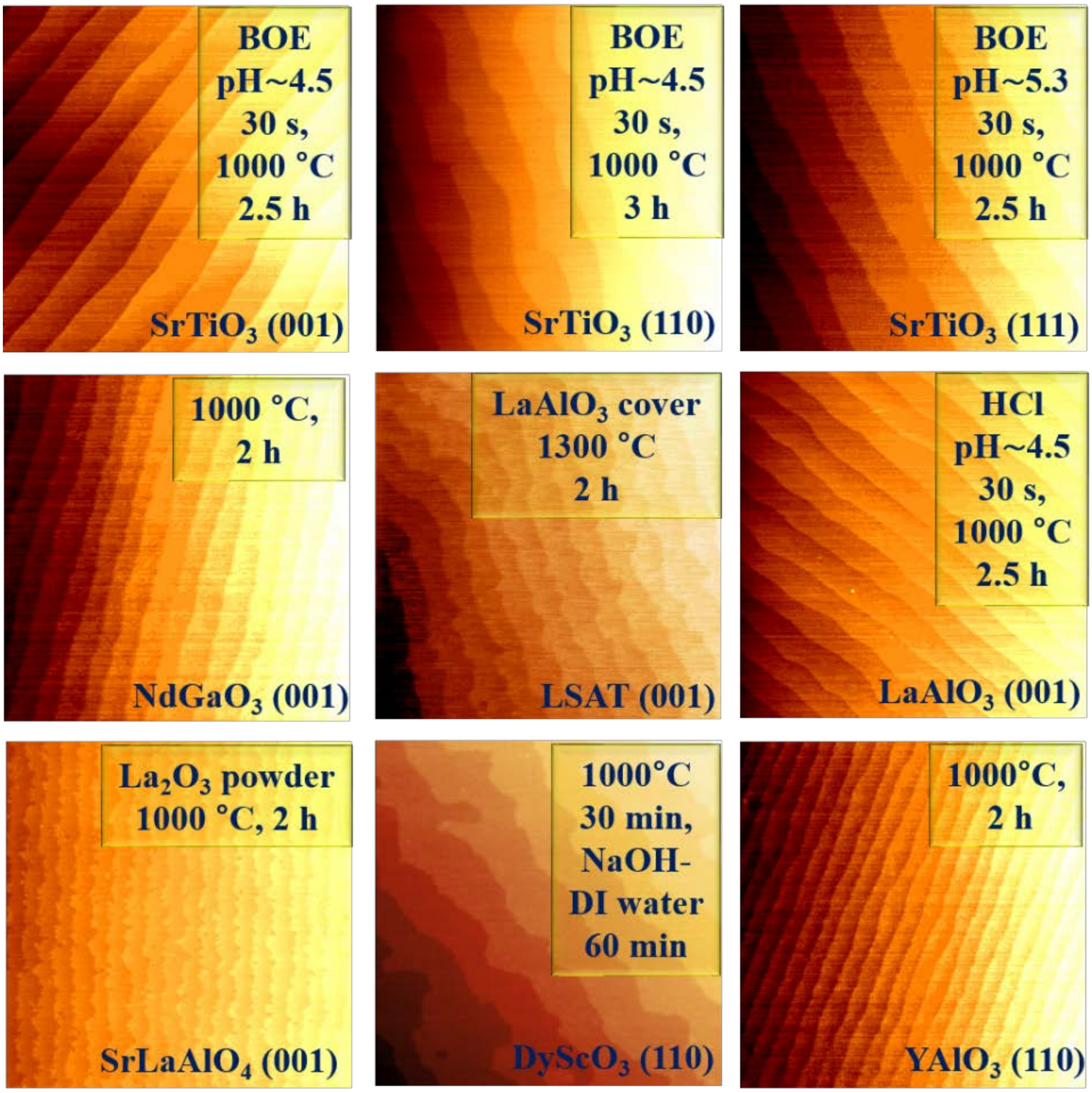

**Fig. 19.** (Color online) Atomically flat surface of various single crystal oxide substrates. Inset shows the treatment conditions for various substrates to obtain atomically flat surfaces. AFM images were either taken from the literature or reproduced by using the methods available in the literature **[80,82,104,147,168,172,173,190,209]**.

**Table 2.** Termination characteristics of treated atomically flat substrate surfaces with lattice constant ranges between 4.00 Å and 3.70 Å.

| Substrate | Orientation | Cubic or pseudo-cubic lattice constant (Å) | Topmost layer termination |
|---|---|---|---|
| $NdScO_3$ | (110) | 4.00 | $ScO_2$ **[168]** |
| $KTaO_3$ | (001) | 3.99 | Not known yet |
| $GdScO_3$ | (110) | 3.96 | $ScO_2$ **[168]** |
| $DyScO_3$ | (110) | 3.94 | $ScO_2$ **[168]** |
| $SrTiO_3$ | (001) | 3.90 | $TiO_2$ **[80]** |
| | (110) | 2.76 | TiO **[104]** |
| | (111) | 2.25 | Ti **[104]** |
| LSAT | (001) | 3.88 | SrO **[82]** |
| $NdGaO_3$ | (001) | 3.86 | NdO **[172]** |
| $LaAlO_3$ | (001) | 3.78 | $AlO_2$ **[172]** |
| $SrLaAlO_4$ | (001) | 3.75 | SrO or $AlO_2$ **[209]** |
| $YAlO_3$ | (110) | 3.72 | Not known yet |

Structurally speaking, metal oxides in general are much more diverse than just perovskites; however, the major focus of oxide thin film research has been on growing perovskite-based materials on top of perovskite substrates as they are isostructural and thus novel interfaces and superlattices are formed coherently without much disorder leading to emergent interfacial phenomena. One of the drawbacks of these epitaxial films and structures is that they usually show the emergent functionalities at low temperatures, limiting a possibility of room temperature device applications. One possible path to be explored for oxide electronics with room temperature functionalities might be the heteroepitaxial synthesis of non-isostructural oxide materials and substrates (e.g. perovskite/spinel, perovskite/hexagonal, perovskite/fluorites, spinel/corundum, and many more) **[217-221]**. For coherent growth of the heterointerfaces, we would need substrates having dimensional and geometrical deviations from perovskites. The candidates for the purpose would include metal oxide single crystals such as spinel $MgAl_2O_4$ ($a$ = 4.04 Å), hexagonal $Al_2O_3$ ($a$ = 4.75 Å), cubic YSZ ($a$ = 5.12 Å), cubic $Gd_3Ga_5O_{12}$ ($a$ = 12.38 Å), and tetragonal $TiO_2$ ($a = b = 4.58$ Å, and $c = 2.95$ Å) **[68,222,223]**. These substrates with atomically flat single terminated topmost layers would allow the exploration of non-isostructural oxide interfaces. We may also mention in passing that surface treatments are important even for measurement purposes as unpolished and non-treated substrates with a trace amount of impurities might cause unusual responses which compete with the responses from ultrathin films. Thus, substrate surface preparation is often crucial in precise determination of the intrinsic magnetic properties of oxide thin films **[224-226]**.

In conclusion, it is our hope that the present collection of oxide substrates and the associated treatment methods would be helpful in practical ways to researchers working in the field of thin film growth. With the rapid progress in the field of interface controlled oxide hetoroepitaxy especially for next generation device applications [**227,228**], one may consider even using atomically flat non-oxide substrate (e.g. $MgF_2$, $a$ = 4.62 Å; and $CaF_2$, $a$ = 5.46 Å) to produce oxide/non-oxide interfaces **[229,230]**. The ability to produce atomically flat surfaces with selective chemical termination would provide an additional degree of freedom in searching for unforeseen emergent phenomena and functional applications in epitaxial oxide thin films and heterostructures with atomically controlled interfaces. Of course, it goes without saying that similar or entirely new procedures should be developed to produce atomically flat single termination on new substrates not covered in this review.

**Acknowledgements**

Authors would like to thank P. B. Rossen for providing us the TOF-MSRI data and Dr. Orlando Auciello for useful comments on setting up the TOF-MSRI instrument. YHJ was supported by National Research Foundation (NRF) of Korea (SRC-2011-0030786 and 2015R1D1A1A02062239). CHY also acknowledges NRF (2014R1A2A2A01005979) for the support.

**Conflict of interest**

The authors declare no competing commercial or financial interest.

**References**

[1] N. Izyumskaya, Ya. Alivov, H. Morkoc, Oxides, Oxides, and More Oxides: High-k Oxides, Ferroelectric, Ferromagnetics, and Multiferroics, Crit. Rev. Solid State Mater. Sci. 34 (2009) 89-179. **(DOI: 10.1080/10408430903368401)**

[2] C. N. R. Rao, B. Raveau, Transition Metal Oxides: Structure, Properties, and Synthesis of Ceramic Oxides, Wiley–VCH, 1998. (ISBN: 978-0-471-18971-8)

[3] S. Maekawa, T. Tohyama, S. E. Barnes, S. Ishihara, W. Koshibae, and G. Khaliullin, Physics of Transition Metal Oxides, Springer Science & Business Media, 2004. (ISBN: 978-3-662-09298-9)

[4] A. P. Ramirez, Colossal magnetoresistance, J. Phys. Codens. Matter 9 (1997) 8171-8199. **(DOI: 10.1088/0953-8984/9/39/005)**

[5] M. Imada, A. Fujimori, Y. Tokura, Metal-insulator transitions, Rev. Mod. Phys. 70 (1998) 103**9. (DOI: 10.1103/RevModPhys.70.1039)**

[6] S. W. Cheong, M. Mostovoy, Multiferroics: a magnetic twist for ferroelectricity, Nat. Mater. 6 (2007) 13-20. **(DOI: 10.1038/nmat1804)**

[7] N. Nagaosa, J. Sinova, A. Onoda, A. H. MacDonald, N. P. Ong, Anomalous Hall effect, Rev. Mod. Phys. 82 (2010) 1539. **(DOI: 10.1103/RevModPhys.82.1539)**

[8] B. Keimer, S. A. Kivelson, M. R. Norman, S. Uchida, J. Zaanen, From quantum matter to high-temperature superconductivity in copper oxides, Nature 518 (2015) 179-186. **(DOI: 10.1038/nature14165)**

[9] W. W. Krempa, G. Chen, Y.-B. Kim, L. Balents, Correlated Quantum Phenomena in the Strong Spin-Orbit Regime, Annu. Rev. Condens. Matter Phys. 5 (2013) 57-82. **(DOI: 10.1146/annurev-conmatphys-020911-125138)**

[10] S. A. Wolf, D. D. Awschalom, R. A. Buhrman, J. M. Daughton, S. von Molnar, M. L. Roukes, A. Y. Chtchelkanova, D. M. Treger, Spintronics: A Spin-Based Electronics Vision for the Future, Science 294 (2001) 1488-1495. **(DOI: 10.1126/science.1065389)**

[11] M. Bibes, A. Barthélémy, Multiferroics: Towards a magnetoelectric memory, Nat. Mater. 7 (2008) 425-426. (**DOI: 10.1038/nmat2189**)

[12] A. Sawa, Resistive switching in transition metal oxides, Mat. Today 11 (2008) 28-36. **(DOI: 10.1016/S1369-7021(08)70119-6)**

[13] L. Fan, B. Zhu, M. Chen, C. Wang, R. Raza, H. Qin, X. Wang, X. Wang. Y. Ma. High performance transition metal oxide composite cathode for low temperature solid oxide fuel cells, J. Power Sources 203 (2012) 65-71. **(DOI: 10.1016/j.jpowsour.2011.12.017)**

[14] X. Yu, T. J. Marks, A. Facchetti, Metal oxides for optoelectronic applications, Nat. Mater. 15 (2016) 383-396. **(DOI: 10.1038/nmat4599)**

[15] G. E. Moore, Cramming More Components onto Integrated Circuits, Electronics 38 (1965) 114. **(DOI: 10.1109/jproc.1998.658762)**

[16] C. Cen, S. Thiel, J. Mannhart, J. Levy, Oxide Nanoelectronics on Demand, Science 323 (2009) 1026-1030. **(DOI: 10.1126/science.1168294)**

[17] J. Mannhart and D. G. Schlom, Oxide Interfaces-An Opportunity for Electronics, Science 327 (2010) 1607-1611. **(DOI: 10.1126/science.1181862)**

[18] D. G. Schlom, J. Mannhart, Oxide electronics: Interface takes charge over Si, Nat. Mater. 10 (2011) 168-169. **(DOI: 10.1038/nmat2965)**

[19] M. Bibes, J. E. Villegas, A. Barthélémy, Ultrathin oxide films and interfaces for electronics and spintronic, Advances in Physics 60 (2011) 5-84. **(DOI: 10.1080/00018732.2010.534865)**

[20] A. S. Bhalla, R. Guo, R. Roy, The perovskite structure- a review of its role in ceramic and technology, Mat. Res. Innovat. 4 (2000) 3-26. **(DOI: 10.1007/s100190000062)**

[21] M. A. Peña, J. L. G. Fierro, Chemical Structures and Performance of Perovskite Oxides, Chem. Rev. 101 (2001) 1981-2018. **(DOI: 10.1021/cr980129f)**

[22] P. K. Davies, H. Wu, A. Y. Borisevich, I. E. Molodetsky, L. Farber, Crystal Chemistry of Complex Perovskites: New Cation-Ordered Dielectric Oxides, Annu. Rev. Mater. Res. 38 (2008) 369-401. **(DOI: 10.1146/annurev.matsci.37.052506.084356)**

[23] L. Pan and G. Zhu, Perovskite Materials - Synthesis, Characterization, Properties, and Applications, Intech, 2016. (ISBN: 978-953-51-2245-6) **(DOI: 10.5772/60469)**

[24] M. Johnsson, P. Lemmens, Perovskites and thin films-crystallography and chemistry, J. Phys.: Condens. Matter. 20 (2008) 264001. **(DOI: 10.1088/0953-8984/20/26/264001)**

[25] L.Q. Jiang, J.K. Guo, H.B. Liu, M. Zhu, X. Zhou, P. Wu, C.H. Li, Prediction of lattice constant in cubic perovskites, J. Phys. Chem. Solids 67 (2006) 1531-1536. **(DOI: 10.1016/j.jpcs.2006.02.004)**

[26] A.S. Verma, V.K. Jindal, Lattice constant of cubic perovskite, J. Alloys and Compd. 485 (2009) 514-518. **(DOI: 10.1016/j.jallcom.2009.06.001)**

[27] A. Kumar, A.S. Verma, Lattice constant of orthorhombic perovskite solids, J. Alloys and Compd. 480 (2009) 650-657. **(DOI: 10.1016/j.jallcom.2009.02.004)**

[28] V. M. Goldschmidt, Die Gesetze der Krystallochemie, Naturwissenschaften 14 (1926) 477-485. **(DOI: 10.1007/BF01507527)**

[29] W. Prellier, Ph. Lecoeur, B. Mercey, Colossal-magnetoresistive manganite thin films, J. Phys. Condens. Matter 13 (2001) R915. **(DOI: 10.1088/0953-8984/13/48/201)**

[30] A-M Haghiri-Gosnet, J-P Renard, CMR manganites: physics, thin films and devices, J. Phys. D: Appl. Phys. 36 (2003) R127. **(DOI: 10.1088/0022-3727/36/8/201)**

[31] D. P. Norton, Synthesis and properties of epitaxial electronic oxide thin-film materials, Mater. Sci. Eng. R. 43 (2004) 139-247. **(DOI: 10.1016/j.mser.2003.12.002)**

[32] M. Dawber, K. M. Rabe, J. F. Scott, Physics of thin-film ferroelectric oxides, Rev. Mod. Phys. 77 (2005) 1083. **(DOI: 10.1103/RevModPhys.77.1083)**

[33] H. N. Lee, H. M. Christen, M. F. Chisholm, C. M. Rouleau, D. H. Lowndes, Strong polarization enhancement in asymmetric three-component ferroelectric superlattices, Nature (London) 433 (2005) 395-399. **(DOI: 10.1038/nature03261)**

[34] R. Ramesh, N. A. Spaldin, Multiferroics: progress and prospects in thin films, Nat. Mater. 6 (2007) 21-29. **(DOI: 10.1038/nmat1805)**

[35] D. G. Schlom, L. -Q. Chen, C. -B. Eom, K. M. Rabe, S. K. Streiffer, J. -M. Triscone, Strain Tuning of Ferroelectric Thin Films, Annu. Rev. Mater. Res. 37 (2007) 589-626. **(DOI: 10.1146/annurev.matsci.37.061206.113016)**

[36] D. G. Schlom, L. -Q. Chen, X. Pan, A. Schmehl, M. A. Zurbuchen, A Thin Film Approach to Engineering Functionality into Oxides, J. Am. Ceram. Soc. 91 (2008) 2429-2454. **(DOI: 10.1111/j.1551-2916.2008.02556.x)**

[37] L. W. Martin, S. P. Crane, Y.- H. Chu, M. B. Holcomb, M. Gajek, M. Huijben, C. –H. Yang, N. Balke, R. Ramesh, Multiferroics and magnetoelectrics: thin films and

nanostructures, J. Phys.: Condens. Matter 20 (2008) 434220. **(DOI: 10.1088/0953-8984/20/43/434220)**

[38] G. Blamire, J. L. MacManus-Driscoll, N. D. Mathur, Z. H. Barber, The Materials Science of Functional Oxide Thin Films, Adv. Mater. 21 (2009) 3827-3839. **(DOI: 10.1002/adma.200900947)**

[39] L. W. Martin, Y. -H. Chu, R. Ramesh, Advances in the growth and characterization of magnetic, ferroelectric, and multiferroic oxide thin films, Mat. Sci. and Eng. R 68 (2010) 89-133. **(DOI: 10.1016/j.mser.2010.03.001)**

[40] L. W. Martin, D. G. Schlom, Advanced synthesis techniques and routes to new single-phase multiferroics, Curr. Opin. Solid State Mater. Sci. 16 (2012) 199-215. **(DOI: 10.1016/j.cossms.2012.03.001)**

[41] M. Gibert, P. Zubko, R. Scherwitzl, J. Íñiguez, J.-M. Triscone, Exchange bias in $LaNiO_3$-$LaMnO_3$ superlattices, Nat. Mater. 11 (2012) 195-198. **(DOI: 10.1038/nmat3224)**

[42] E. J. Monkman, C. Adamo, J. A. Mundy, D. E. Shai, J. W. Harter, D. Shen, B. Burganov, D. A. Muller, D. G. Schlom, K. M. Shen, Quantum many-body interaction in digital oxide superlattices, Nat. Mater. 11 (2012) 855-859. **(DOI: 10.1038/nmat3405)**

[43] M. Dawber, E. Bousquet, New developments in artificially layered ferroelectric oxide superlattics, MRS Bull. 38 (2013) 1048-1055**. (DOI: 10.1557/mrs.2013.263)**

[44] A. Bhattacharya, S. J. May, Magnetic Oxide Heterostrcutures, Annu. Rev. Mater. Res. 44 (2014) 65-90. **(DOI: 10.1146/annurev-matsci-070813-113447)**

[45] D. G. Schlom, L.-Q. Chen, C. J. Fennie, V. Gopalan, D. A. Muller, X. Pan, R. Ramesh, R. Uecker, Elastic strain engineering of ferroic oxides, MRS Bull. 39 (2014) 118-130. **(DOI: doi.org/10.1557/mrs.2014.1)**

[46] J. T. Heron, D. G. Schlom, R. Ramesh, Electric field control of magnetism using $BiFeO_3$-based heterostructures, Appl. Phys. Rev. 1 (2014) 021303. **(DOI: 10.1063/1.4870957)**

[47] J. A. Mundy, C. M. Brooks, M. E. Holtz, J. A. Moyer, H. Das, A. F. Rebola, J. T. Heron, J. D. Clarkson, S. M. Disseler, Z. Liu, A. Farhan, R. Held, R. Hovden, E. Padgett, Q. Mao, H. Piak, R. Mishra, L. F. Koukoutis, E. Arenholz, A. Scholl, J. A.

Borhers, W. D. Ratcliff, R. Ramesh, C. J. Fennie, P. Schiffer, D. A. Muller, D. G. Schlom, Atomically engineered ferroic layers yield a room-temperature magnetoelectric multiferroic, Nature 537 (2016) 523-527. **(DOI: 10.1038/nature19343)**

[48] A. R. Damodaran, J. C. Agar, S. Pandya, Z. Chen, L. Dedon, R. Xu, B. Apgar, S. Saremi, L. W. Martin, New modalities of strain-control of ferroelectric thin films, J. Phys.: Condens. Matter 28 (2016) 263001. **(DOI: 10.1088/0953-8984/28/26/263001)**

[49] H. Boschker, J. Mannhart, Quantum-Matter Heterostructures, Annu. Rev. Condens. Matter Phys. 8 (2017) 145. **(DOI: 10.1146/annurev-conmatphys-031016-025404)**

[50] A. Ohtomo, H. Y. Hwang, A high-mobility electron gas at the $LaAlO_3$/$SrTiO_3$ heterointerface, Nature (London) 427 (2004) 423-426. **(DOI: 10.1038/nature02308)**

[51] J. Mannhart, D. H. A. Blank, H. Y. Hwang, A. J. Millis, J. –M. Triscone, Two-Dimensional Electron Gases at Oxide Interfaces, MRS Bull. 33 (2008) 1027-1034. **(DOI: 10.1557/mrs2008.222)**

[52] A. Brinkman, M. Huijben, M. van Zalk, J. Huijben, U. Zeitler, J. C. Maan, W. G. van der Wiel, G. Rijnders, D. H. A. Blank, H. Hilgenkamp, Magnetic effects at the interface between non-magnetic oxides, Nat. Mater. 6 (2007) 493-496. **(DOI: 10.1038/nmat1931)**

[53] N. Reyren, S. Thiel, A. D. Caviglia, L. Fitting Kourkoutis, G. Hammerl, C. Richter, C. W. Schneider, T. Kopp, A.-S. Ruetschi, D. Jaccard, M. Gabay, D. A. Muller, J.-M. Triscone, J. Mannhart, Superconducting Interfaces Between Insulating Oxides, Science 317 (2007) 1196-1199. **(DOI: 10.1126/science.1146006)**

[54] M. Huijben, A. Brinkman, G. Koster, G. Rijnders, H. Hilgenkamp, D. H. A. Blank, Structure-Property Relation of $SrTiO_3$/$LaAlO_3$ Interfaces, Adv. Mater. 21 (2009) 1665-1677. **(DOI: 10.1002/adma.200801448)**

[55] P. Zubko, S. Gariglio, M. Gabay, P. Ghosez, J.-M. Triscone, Interface Physics in Complex Oxide Heterostructures, Annu. Rev. Condens. Matter Phys. 2 (2011) 141-165. **(DOI: 10.1146/annurev-conmatphys-062910-140445)**

[56] H. Y. Hwang, Y. Iwasa, M. Kawasaki, B. Keimer, N. Nagaosa, Y. Tokura, Emergent phenomena at oxide interfaces, Nat. Mater. 11 (2012) 103-113. **(DOI: 10.1038/nmat3223)**

[57] P. Yu, Y. –H. Chu, R. Ramesh, Oxide interfaces: pathways to novel phenomena, Mat. Today 15 (2012) 320-327. **(DOI: 10.1016/S1369-7021(12)70137-2)**

[58] F. M. Granozio, G. Koster, G. Rijnders, Functional oxide interfaces, MRS Bull. 38 (2013) 1017-1023. **(DOI: 10.1557/mrs.2013.282)**

[59] J. Chakhalian, J. W. Freeland, A. J. Millis, Panagopoulos, J. M. Rondinelli, *Colloquium*: Emergent properties in plane view: Strong correlations at oxide interfaces, Rev. Mod. Phys. 86 (2014) 1189. **(DOI: 10.1103/RevModPhys.86.1189)**

[60] S. Stemmer, S. J. Allen, Two-Dimensional Electron Gases at Complex Oxide Interfaces, Annu. Rev. Mater. Res. 44 (2014) 151-171. **(DOI: 10.1146/annurev-matsci-070813-113552)**

[61] D. Dijkkamp, T. Venkatesan, X. D. Wu, S. A. Shaheen, N. Jisrawi, Y. H. Min-Lee, W. L. Mclean, M. Croft, Preparation of Y-Ba-Cu oxide superconductor thin films using pulsed laser evaporation from high $T_c$ bulk material, Appl. Phys. Lett. 51 (1987) 619. **(DOI: 10.1063/1.98366)**

[62] D. B. Chrisey, G. K. Hubler, Pulsed Laser Deposition of Thin Films, John Wiley & Sons, 1994. (ISBN: 0-471-59218-8)

[63] S. B. Ogale, Thin Films and Heterostructures for Oxide Electronics, Springer US, 2005. (ISBN: 978-0-387-26089-1)

[64] H. M. Christen, G. Eres, Recent advances in pulsed-laser deposition of complex oxides, J. Phys.: Condens. Matter 20 (2008) 264005. **(DOI: 10.1088/0953-8984/20/26/264005)**

[65] M. Opel, Spintronic oxide grown by laser-MBE, J. Phys. D: Appl. Phys. 45 (2012) 033001. **(DOI: 10.1088/0022-3727/45/3/033001)**

[66] G. Koster, M. Huijben, G. Rijnders, Epitaxial Growth of Complex Metal Oxides, Elsevier Science & Technology, 2015. (ISBN: 978-1-78242-245-7)

[67] J. M. Phillips, Substrate selection for high-temperature superconducting thin films, J. Appl. Phys. 79 (1996) 1829. **(DOI: 10.1063/1.362675)**

[68] CrysTec, Germany (http://www.crystec.de/products-e.html);

MTI, USA (http://www.mtixtl.com/crystalssubstratesa-z.aspx);

Shinkosha, Japan (http://www.shinkosha.com/english/sehin/index.html).

[69] N. Nakagawa, H. Y. Hwang, D. A. Muller, Why some interfaces cannot be sharp, Nat. Mater. 5 (2006) 204. **(DOI: 10.1038/nmat1569)**

[70] L. Wang, K. F. Ferris, G. S. Herman, Interactions of $H_2O$ with $SrTiO_3$ (100) surfaces, J. Vac. Sci. Technol. A, 20 (2002) 239. **(DOI: 10.1116/1.1430246)**

[71] J. M. P. Martirez, S. Kim, E. H. Morales, B. T. Diroll, M. Cargnello, T. R. Gordon, C. B. Murray, D. A. Bonnell, A. M. Rappe, Synergistic Oxygen Evolving Activity of a $TiO_2$-Rich Reconstructed $SrTiO_3$ (001) surface, J. Am. Chem. Soc. 137 (2015) 2939. **(DOI: 10.1021/ja511332y)**

[72] R. Pentcheva, W. E. Pickett, Electronic phenomena at complex oxide interfaces: insights from first principles, J. Phys: Condens. Matter 22 (2010) 043001. **(DOI: 10.1088/0953-8984/22/4/043001)**

[73] N. C. Bristowe, P. Ghosez, P. B. Littlewood, E. Artacho, The origin of two-dimensional electron gases at oxide interfaces: insights from theory, J. Phys.: Condens Matter 26 (2014) 143201. **(DOI: 10.1088/0953-8984/26/14/143201)**

[74] B. Kim, B. I. Min, Termination-dependent electronic and magnetic properties of ultrathin $SrRuO_3$ (111) films on $SrTiO_3$. Phys. Rev. B 89 (2014) 195411. **(DOI: 10.1103/PhysRevB.89.195411)**

[75] S. Nazir, J. Cheng, K. Yang, Creating Two-dimensional Electron Gas in Nonpolar/Nonpolar Oxide Interface via Polarization Discontinuity: Fist-Principle Analysis of $CaZrO_3$/$SrTiO_3$ Heterostructure, ACS Appl. Mater. Interfaces 8 (2016) 390. **(DOI: 10.1021/acsami.5b09107)**

[76] V. E. Henrich, METAL-OXIDE SURFACES, Prog. Surf. Sci. 50 (1995) 77-90. **(DOI: 10.1016/0079-6816(95)00046-1)**

[77] C. Noguera, Physics and Chemistry at Oxide Surfaces, Cambridge University Press, 1996. (ISBN: 0-52147214-8).

[78] C. Noguera, Polar oxide surfaces, J. Phys.: Condens. Matter 12 (2000) R367. **(DOI: 10.1088/0953-8984/12/31/201)**

[79] J. Goniakowski, F. Finocchi, C. Noguera, Polarity of oxide surfaces and nanostructures, Rep. Prog. Phys. 71 (2008) 016501. **(DOI: 10.1088/0034-4885/71/1/016501)**

[80] M. Kawasaki, K. Takahashi, T. Maeda, R. Tsuchiya, M. Shinohada, O. Ishiyama, T. Yonezawa, M. Yoshimoto, H. Koinuma, Atomic Control of the $SrTiO_3$ Crystal Surface, Science 266 (1994) 1540-1542. **(DOI: 10.1126/science.266.5190.1540)**

[81] J. G. Connell, B. J. Isaac, G. B. Ekanayake, D. R. Strachan, S. S. A. Seo, Preparation of atomically flat $SrTiO_3$ surfaces using a deionized-water leaching and thermal annealing procedure, Appl. Phys. Lett. 101 (2012) 251607. **(DOI: 10.1063/1.4773052)**

[82] J. H. Ngai, T. C. Schwendemann, A. E. Walker, Y. Segal, F. J. Walker, E. I. Altman, C. H. Ahn, Achieving A-Site Termination on $La_{0.18}Sr_{0.82}Al_{0.59}Ta_{0.41}O_3$ Substrates, Adv. Mater. 22 (2010) 2945-2978. **(DOI: 10.1002/adma.200904328)**

[83] J. Wayne Rabalais, Principle and Applications of Ions Scattering Spectroscopy: Surface Chemical and Structural Analysis, Wiley NJ, 2003. (ISBN: 978-0-471-20277-6).

[84] O. Auciello, A. R. Krauss, J. Im, J. A. Schultz, STUDIES OF MULTICOMPONENT OXIDE FILMS AND LAYERED HETEROSTRCUTURE GROWTH PROCESSES VIA IN SITU, TIME-OF-FLIGHT ION SCATTERING AND DIRECT RECOIL SPECTROSCOPY, Annu. Rev. Mater. Sci. 28 (1998) 375-296. **(DOI: 10.1146/annurev.matsci.28.1.375)**

[85] R. J. Cannara, M. Eglin, R. W. Carpick, Lateral force calibration in atomic force microscopy: A new calibration method and general guidelines for optimization, Rev. Sci, Instrum 77 (2006) 053701. **(DOI: 10.1063/1.2198768)**

[86] G. Binnig, H. Rohrer, Scanning tunneling microscopy-from birth to adolescence, Rev. Mod. Phys. 59 (1987) 615. **(DOI: 10.1103/RevModPhys.59.615)**

[87] J. P. Podkaminer, J. J. Patzner, B. A. Davidson, C. B. Eom, Real-time in situ monitoring of sputter deposition with RHEED for atomic layer controlled growth, APL Mater. 4 (2016) 086111. **(DOI: 10.1063/1.4961503)**

[88] F. Jona, J. A. Strozier, W. S. Yang, Low-energy electron diffraction for surface structure analysis, Rep. Prog. Phys. 45 (1982) 527. **(DOI: 10.1088/0034-4885/45/5/002)**

[89] D. C. Koningsberger, R. Prins, X-ray absorption: principle, applications, techniques of EXAFS, SEXAFS, AND XANES, John Wiley and Sons, New York, NY, 1988. (ISBN: 978-0-471-87547-5)

[90] S. Hüfner, Photoelectron Spectroscopy-Principles and Applications, Springer-Verlag, Berlin, 2003. (ISBN: 978-3-662-09280-4)

[91] K. van Benthem, C. Elsässer, R. H. French, Bulk electronic structure of $SrTiO_3$: Experiment and theory, J. Appl. Phys. 90 (2001) 6156. **(DOI: 10.1063/1.1415766)**

[92] G. Herranz, F. Sanchez, N. Dix, M. Scigaj, J. Fontcuberta, High mobility conduction at (110) and (111) $LaAlO_3/SrTiO_3$ interfaces, Sci. Rep. 2 (2012) 758. **(DOI: 10.1038/srep00758)**

[93] A. Annadi, Q. Zhang, X. R. Wang, N. Tuzia, K. Gopinadhan, W. M. Lu, A. R. Barman, Z. Q. Liu, A. Srivastava, S. Saha, Y. L. Zhao, S. W. Zeng, S. Dhar, E. Olsson, B. Gu, S. Yunoki, S. Maekawa, H. Hilgenkamp, T. Venkatesan, Ariando, Anisotropic two-dimensional electron gas at the $LaAlO_3/SrTiO_3$ (110) interface, Nat. Commun. 4 (2013) 1838. **(DOI: 10.1038/ncomms2804)**

[94] A. F. Santerder-Syro, O. Copie, T. Kondo, F. Fortuna, S. Pailhes, R. Weht, X. G. Qiu, F. Bertran, A. Nicolaou, A. Taleb-Ibrahimi, P. Le Fevre, G. Herranz, M. Bibes, N. Reyren, Y. Apertet, P. Lecoeur, A. Barthelemy, M. J. Rozenderg, Two-dimensional electron gas with universal subbands at the surface of $SrTiO_3$, Nature 469 (2011) 189-193. **(DOI: 10.1038/nature09720)**

[95] W. Meevasana, P. D. C. King, R. H. He, S. –K. Mo, M. Hashimoto, A. Tamai, P. Songsiriritthigul, F. Baumberger, Z. –X. Chen, Creation and control of a two-dimensional electron liquid at the bare $SrTiO_3$ surface, Nat. Mater. 10 (2011) 114-118. **(DOI: 10.1038/nmat2943)**

[96] F. Sánchez, C. Ocal, J. Fontcuberta, Tailored surfaces of perovskite oxide substrates for conducted growth of thin films, Chem. Soc. Rev. 43 (2014) 2272-2285. **(DOI: 10.1039/C3CS60434A)**

[97] M. Kawasaki, A. Ohtomo, T. Arakane, K. Takahashi, M. Yoshimoto, H. Koinuma, Atomic control of $SrTiO_3$ surface for perfect epitaxy of perovskite oxides, Appl. Surf. Sci. 107 (1996) 102-106. **(DOI: 10.1016/S0169-4332(96)00512-0)**

[98] G. Koster, B. L. Kropman, G. J. H. M. Rijnders, D. H. A. Blank, H. Rogalla, Quasi-ideal strontium titanate crystal surfaces through formation of strontium hydroxide, Appl. Phys. Lett. 73 (1998) 2920. **(DOI: 10.1063/1.122630)**

[99] G. Koster, G. Rijnders, D. H. A. Blank, H. Rogalla, Surface morphology determined by (001) single-crystal $SrTiO_3$ termination, Physica C 339 (2000) 215-230. **(DOI: 10.1016/S0921-4534(00)00363-4)**

[100] D. Kobyashi, H. Kumigashira, M. Oshima, T. Ohnishi, M. Lippmaa, K. Ono, M. Kawasaki, H. Koinuma, High-resolution synchrotron-radiation photoemission characterization for atomically-controlled $SrTiO_3$ (001) substrate surfaces subjected to various surface treatments, J. App. Phys. 96 (2004) 7183. **(DOI: 10.1063/1.1814175)**

[101] T. Ohnishi, K. Shibuya, M. Lippmaa, D. Kobayashi, H. Koinuma, M. Oshima, H. Koinuma, Preparation of thermally stable $TiO_2$-terminated $SrTiO_3$ (100) substrate surfaces, Appl. Phys. Lett. 85 (2004) 272. **(DOI: 10.1063/1.1771461)**

[102] A. Fragneto, G. M. De Luca, R. Di Capua, U. Scotti di Uccio, M. Salluzzo, X. Torrelles, Tien-Lin Lee, J. Zegenhagen, Ti- and Sr-rich surfaces of $SrTiO_3$ studied by grazing incidence x-ray diffraction, Appl. Phys. Lett. 91 (2007) 101910. **(DOI: 10.1063/1.2779972)**

[103] R. Gunnarssson, A. S. Kalabukhov, D. Winkler, Evaluation of recipes for obtaining single terminated perovskite oxide substrates, Surf. Sci. 603 (2009) 151-157. **(DOI: 10.1016/j.susc.2008.10.045)**

[104] A. Biswas, P. B. Rossen, C.-H. Yang, W. Siemons, M.-H. Jung, I. K. Yang, R. Ramesh, Y. H. Jeong, Universal Ti-rich termination of atomically flat $SrTiO_3$ (001), (110), and (111) surfaces, Appl. Phys. Lett. 98 (2011) 051904. **(DOI: 10.1063/1.3549860)**

[105] C. Raisch, T. Chassé, Ch. Langheinrich, A. Chassé, Preparation and investigation of the A-site and B-site terminated $SrTiO_3$ (001) surface: A combined experimental and theoretical x-ray photoelectron diffraction study, J. Appl. Phys. 112 (2012) 073505. **(DOI: 10.1063/1.4757283)**

[106] S. A. Chambers, T. C. Droubay, C. Capan, G. Y. Sun, Unintentional F doping of $SrTiO_3$ (001) etched in HF acid-structure and electronic properties, Sur. Sci. 606 (2012) 554-558. **(DOI: 10.1016/j.susc.2011.11.029)**

[107] V. Leca, G. Rijnders, G. Koster, D. H. A. Blank, H. Rogalla, Wet Etching Methods for Perovskite Substrates, Mat. Res. Soc. Symp. 587 (2000) O3. 6. 1. **(DOI: 10.1557/PROC-587-O3.6)**

[108] J. Zhang, D. Doutt, T. Merz, J. Chakhalian, M. Kareev, J. Liu, L. J. Brillson, Depth-resolved subsurface defects in chemically etched $SrTiO_3$, Appl. Phys. Lett. 94 (2009) 092904. **(DOI: 10.1063/1.3093671)**

[109] I. V. Davalos, R. Thomas, A. Ruediger, Realization of single-termination $SrTiO_3$ (100) surfaces by a microwave-induced hydrothermal process, Appl. Phys. Lett. 103 (2013) 202905. **(DOI: 10.1063/1.4831681)**

[110] R. C. Hatch, K. D. Fredrickson, M. Choi, C. Lin, H. Seo, A. B. Posadas, A. A. Demkov, Surface electronic structure for various surface preparations of Nb-doped $SrTiO_3$ (001), J. Appl. Phys. 114 (2013) 103710. **(DOI: 10.1063/1.4821095)**

[111] R. C. Hatch, M. Choi, A. B. Posadas, A. A. Demkov, Comparison of acid- and non-acid-based surface preparations of Nb-doped $SrTiO_3$ (001), J. Vac. Sci. Technol. B 33 (2015) 061204. **(DOI: 10.1116/1.4931616)**

[112] R. Bachelet, F. Sánchez, F. J. Palomares, C. Ocal, J. Fontcuberta, Atomically flat SrO-terminated $SrTiO_3$ (001) substrate, Appl. Phys. Lett. 95 (2009) 141915. **(DOI: 10.1063/1.3240869)**

[113] R. Bachelet, F. Sa´nchez, J. Santiso, C. Munuera, C. Ocal, J. Fontcuberta, Self-Assembly of $SrTiO_3$ (001) Chemical-Terminations: A Route for Oxide Nanostructure Fabrication by Selective Growth, Chem. Mater. 21 (2009) 2494-2498. **(DOI: 10.1021/cm900540z)**

[114] K. Szot, W. Speier, Surface of reduced and oxidized $SrTiO_3$ from atomic force microscopy, Phys. Rev. B 60 (1999) 5909. **(DOI: 10.1103/PhysRevB.60.5909)**

[115] K. Szot, W. Speier, U. Breuer, R. Meyer, J. Szade, R. Waser, Formation of micro-crystals on the (100) surface of $SrTiO_3$ at elevated temperatures, Surf. Sci. 460 (2000) 112-128. **(DOI: 10.1016/S0039-6028(00)00522-7)**

[116] H. Wei, L. Beuermann, J. Helmbold, G. Borchardt, V. Kempter, G. Lilienkamp, W. M-Friedrichs, Study of SrO segregation on $SrTiO_3$ (100) surfaces, J. Eur. Ceram. Soc. 21 (2001) 1677-1680. **(DOI: 10.1016/S0955-2219(01)00091-7)**

[117] T. Matsumoto, H. Tanaka, T. Kawai, S. Kawai, STM-imaging of a $SrTiO_3$ (100) surface with atomic-scale resolution, Surf. Sci. Lett. 278 (1992) L153-L158. **(DOI: 10.1016/0167-2584(92)90249-5)**

[118] H. Tanaka, T. Matsumoto, T. Kawai, S. Kawai, Interaction of oxygen vacancies with $O_2$ on a reduced $SrTiO_3$ (100)$\sqrt{5} \times \sqrt{5}$-$R$26.6° surface observed by STM, Surf. Sci. 318 (1994) 29-38. **(DOI: 10.1016/0039-6028(94)90338-7)**

[119] T. Kubo, H. Nozoye, Surface Structure of $SrTiO_3$ (100)-($\sqrt{5} \times \sqrt{5}$)-$R$26.6°, Phys. Rev. Lett. 86 (2001) 1801. **(DOI: 10.1103/PhysRevLett.86.1801)**

[120] Q. D. Jiang, J. Zegenhagen, c(6×2) and c(4×2) reconstruction of $SrTiO_3$ (001), Surf. Sci. 425 (1999) 343-354. **(DOI: 10.1016/S0039-6028(99)00223-X)**

[121] M. R. Castell, Nanostructures on the $SrTiO_3$ (001) surface studied by STM, Surf. Sci. 516 (2002) 33-42. **(DOI: 10.1016/S0039-6028(02)02053-8)**

[122] N. Erdman, K. R. Poeppelmeier, M. Asta, O. Warschkow, D. E. Ellis, L. D. Marks, The structure and chemistry of the $TiO_2$-rich surface of $SrTiO_3$ (001), Nature 419 (2002) 55-58. **(DOI: 10.1038/nature01010)**

[123] N. Erdman, O. Warschkow, M. Asta, K. R. Poeppelmeier, D. E. Ellis, L. D. Marks, Surface Structures of $SrTiO_3$ (001): A $TiO_2$-rich Reconstruction with a c(4×2) Unit Cell, J. Am. Chem. Soc. 125 (2003) 10050-10056. **(DOI: 10.1021/ja034933h)**

[124] F. Silly, D. T. Newell, M. R. Castell, $SrTiO_3$ (001) reconstructions; the (2×2) to c(4×4) transition, Surf. Sci. 600 (2006) 219-223. **(DOI: 10.1016/j.susc.2006.05.043)**

[125] R. Herger, P. R. Willmott, O. Bunk, C. M. Schleputz, B. D. Patterson, B. Delley, V. L. Shneerson, P. F. Lyman, D. K. Saldin, Surface structure of $SrTiO_3$ (001), Phys. Rev. B 76 (2007) 195435. **(DOI: 10.1103/PhysRevB.76.195435)**

[126] R. Herger, P. R. Willmott, O. Bunk, C. M. Schleputz, B. D. Patterson, B. Delley, Surface of Strontium Titanate, Phys. Rev. Lett. 98 (2007) 076102. **(DOI: 10.1103/PhysRevLett.98.076102)**

[127] Y. Lin, A. E. Becerra-Toledo, F. Silly, K. R. Poeppelmeier, M. R. Castell, L. D. Marks, The (2×2) reconstructions on the $SrTiO_3$ (001) surface: A combined scanning tunneling microscopy and density functional theory study, Surf. Sci. 605 (2011) L51-L55. **(DOI: 10.1016/j.susc.2011.06.001)**

[128] O. E. Dagdeviren, G. H. Simon, K. Zou, F. J. Walker, C. Ahn, E. I. Altman, U. D. Schwarz, Surface phase, morphology, and charge distribution transitions on vacuum and ambient annealed $SrTiO_3$ (100), Phys. Rev. B 93 (2016) 195303. **(DOI: 10.1103/PhysRevB.93.195303)**

[129] A. E. Becerra-Toledo, M. R. Castell, L. D. Marks, Wet adsorption on $SrTiO_3$(001): I. Experimental and simulated STM, Surf. Sci. 606 (2012) 762-765. **(DOI: 10.1016/j.susc.2012.01.008)**

[130] A. E. Becerra-Toledo, J. A. Enterkin, D. M. Kienzle, L. D. Marks, Water adsorption on $SrTiO_3$ (001): II. Water, water, everywhere, Surf. Sci. 606 (2012) 791-802. **(DOI: 10.1016/j.susc.2012.01.010)**

[131] J. Padilla, D. Vanderbilt, Ab intio study of $SrTiO_3$ surfaces, Surf. Sci. 418 (1998) 64-70. **(DOI: 10.1016/S0039-6028(98)00670-0)**

[132] O. Warschkow, M. Asta, N. Erdman, K. R. Poeppelmeier, D. E. Ellis, L. D. Marks, $TiO_2$-rich reconstructions of $SrTiO_3$ (001): a theoretical study of structural patterns, Surf. Sci. 573 (2004) 446-456. (**DOI: 10.1016/j.susc.2004.10.012)**

[133] J. Wang, M. Fu, X. S. Wu, D. Bai, Surface structure of strontium titanate, J. Appl. Phys. 105 (2009) 083526. **(DOI: 10.1063/1.3106615)**

[134] G. Z. Zhu, G. Radtke, G. A. Botton, Bonding and structure of a reconstructed (001) surface of $SrTiO_3$ from TEM, Nature 490 (2012) 384-387. **(DOI: 10.1038/nature11563)**

[135] Y. Mukunoki, N. Nakagawa, T. Susaki, H. Y. Hwang, Atomically flat (110) $SrTiO_3$ and heteroepitaxy, Appl. Phys. Lett. 86 (2005) 171908. **(DOI: 10.1063/1.1920415)**

[136] A. Pojani, F. Finocchi, C. Noguera, Polarity on the $SrTiO_3$ (111) and (110) surfaces, Surf. Sci. 442 (1999) 179-198. **(DOI: 10.1016/S0039-6028(99)00911-5)**

[137] E. Heifets, W. A. Goddard III, E. A. Kotomin, R. I. Eglitis, G. Borstel, Ab initio calculations of the $SrTiO_3$ (110) polar surface, Phys. Rev. B 69 (2004) 035408. **(DOI: 10.1103/PhysRevB.69.035408)**

[138] F. Bottin, F. Finocchi, C. Noguera, Facetting and ($n \times 1$) reconstructions of $SrTiO_3$ (110) surfaces, Surf. Sci. 574 (2005) 65-76. **(DOI: 10.1016/j.susc.2004.10.037)**

[139] R. Bachelet, F. Valle, I. C. Infante, F. Sánchez, J. Fontcuberta, Step formation, faceting, and bunching in atomically flat $SrTiO_3$ (110) surfaces, Appl. Phys. Lett. 91 (2007) 251904. **(DOI: 10.1063/1.2825586)**

[140] H. Bando, Y. Aiura, Y. Haruyama, T. Shimizu, Y. Nishihara, Structure and electronic states on reduced $SrTiO_3$ (110) surface observed by scanning tunneling microscopy and spectroscopy, J. Vac. Sci. Technol. B 13 (1995) 1150. **(DOI: 10.1116/1.588227)**

[141] Z. Wang, F. Yang, Z. Zhang, Y. Tang, J. Feng, K. Wu, Q. Guo, J. Guo, Evaluation of the surface structures on $SrTiO_3$ (110) tuned by Ti or Sr concentration, Phys. Rev. B 83 (2011) 155453. **(DOI: 10.1103/PhysRevB.83.155453)**

[142] J. Brunen, J. Zegenhagen, Investigation of the $SrTiO_3$ (110) surface by means of LEED, scanning tunneling microscopy and Auger spectroscopy, Surf. Sci. 389 (1997) 349-365. **(DOI: 10.1016/S0039-6028(97)00450-0)**

[143] B. C. Russell, M. R. Castell, Reconstructions on the polar $SrTiO_3$ (110) surface: Analysis using STM, LEED, and AES, Phys. Rev. B 77 (2008) 245414. **(DOI: 10.1103/PhysRevB.77.245414)**

[144] J. A. Enterkin, A. K. Subramanian, B. C. Russell, M. R. Castell, K. R. Poeppelmeier, L. D. Marks, A homologous series of structures on the surface of $SrTiO_3$ (110), Nat. Mater. 9 (2010) 245-248. **(DOI: 10.1038/nmat2636)**

[145] H. Tanaka, T. Kawai, Surface structure of reduced $SrTiO_3$ (111) observed by scanning tunneling microscopy, Surf. Sci. 365 (1996) 437-442. **(DOI: 10.1016/0039-6028(96)00739-X)**

[146] S. Sekiguchi, M. Fujimoto, M. Nomura, S.-B Cho, J. Tanaka, T. Nishihara, M.-G. Kang, H. –H. Park, Atomic force microscopic observation of $SrTiO_3$ polar surface, Solid State Ionics 108 (1998) 73-79. **(DOI: 10.1016/S0167-2738(98)00021-6)**

[147] J. Chang, Y. -S. Park, S. -K. Kim, Atomically flat single-terminated $SrTiO_3$ (111) surface, Appl. Phys. Lett. 92 (2008) 152910. **(DOI: 10.1063/1.2913005)**

[148] J. L. Blok, X. Wan, G. Koster, D. H. A. Blank, G. Rijnders, Epitaxial oxide growth on polar (111) surfaces, Appl. Phys. Lett. 99 (2011) 151917. **(DOI: 10.1063/1.3652701)**

[149] M. Saghayezhian, L. Chen, G. Wang, H. Guo, E. W. Plummer, J. Zhang, Polar compensation at the surface of $SrTiO_3$ (111), Phys. Rev. B 93 (2016) 125408. **(DOI: 10.1103/PhysRevB.93.125408)**

[150] T. –D. Doan, J. L. Giocondi, G. S. Rohrer, P. A. Salvador, Surface engineering along the close-packed direction of $SrTiO_3$, J. Crys. Growth 225 (2001) 178-182. **(DOI: 10.1016/S0022-0248(01)00829-6)**

[151] B. C. Russell, M. R. Castell, $(\sqrt{13}\times\sqrt{13})R13.9°$ and $(\sqrt{7}\times\sqrt{7})R19.1°$ reconstructions of the polar $SrTiO_3$ (111) surface, Phys. Rev. B 75 (2007) 155433. **(DOI: 10.1103/PhysRevB.75.155433)**

[152] B. C. Russell, M. R. Castell, Surface of Sputtered and Annealed Polar $SrTiO_3$ (111): $TiO_x$-Rich ($n$×$n$) Reconstructions, J. Phys. Chem. C 112 (2008) 6538-6545. **(DOI: 10.1021/jp711239t)**

[153] L. D. Marks, A. N. Chiaramonti, F. Tran, P. Blaha, The small unit cell reconstructions of $SrTiO_3$ (111), Surf. Sci. 603 (2009) 2179-2187. **(DOI: 10.1016/j.susc.2009.04.016)**

[154] L. D. Marks, A. N. Chiaramonti, S. U. Rahman, M. R. Castell, Transition from Order to Configurational Disorder for Surface Reconstructions on $SrTiO_3$ (111), Phys. Rev. Lett. 114 (2015) 226101. **(DOI: 10.1103/PhysRevLett.114.226101)**

[155] S. Woo, H. Jeong, S. A Lee, H. Seo, M. Lacotte, A. David, H. Y. Kim, W. Prellier, Y. Kim, W. S. Choi, Surface properties of atomically flat polycrystalline $SrTiO_3$, Sci. Rep. 5 (2015) 8822. **(DOI: 10.1038/srep08822**)

[156] J. Nie, A. Shoji, M. Koyanagi, H. Takashima, N. Terada, K. Endo, Control of Step Arrays on Normal and Vicinal $SrTiO_3$ (100) Substrates, Jpn. J. Appl. Phys. 37 (1998) L1014. **(DOI: 10.1143/JJAP.37.L1014)**

[157] J. C. Jiang, W. Tian, X. Pan, Q. Gan, C. B. Eom, Effects of miscut of the $SrTiO_3$ substrate on microstructures of the epitaxial $SrRuO_3$ thin films, Mater. Sci. and Eng., B 56 (1998) 152-157. **(DOI: 10.1016/S0921-5107(98)00227-X)**

[158] F. Sánchez, G. Herranz, I. C. Infante, J. Fontcuberta, M. V. García-Cuenca, C. Ferrater, M. Varela, Critical effects of substrate terraces and steps morphology on the growth mode of epitaxial $SrRuO_3$ films, Appl. Phys. Lett. 85 (2004) 1981. **(DOI: 10.1063/1.1786361)**

[159] W. Hong, H. N. Lee, M. Soon, H. M. Christen, D. H. Lowndes, Z. Suo, Z. Zhang, Persistent Step-Flow Growth of Strained Films on Vicinal Substrates, Phys. Rev. Lett. 95 (2005) 095501. **(DOI: 10.1103/PhysRevLett.95.095501)**

[160] E. D Williams, Surface steps and surface morphology: understanding macroscopic phenomena from atomic observations, Surf. Sci. 299/300 (1994) 502-524. **(DOI: 10.1016/0039-6028(94)90678-5)**

[161] G. B. Cho, Y. Kamada, M. Yamamoto, Studies of Self-Organized Steps and Terraces in inclined $SrTiO_3$ (001) Substrate by Atomic Force Microscopy, Jpn. J. Appl. Phys. 40 (2001) 4666. **(DOI: 10.1143/JJAP.40.4666)**

[162] A. Pimpinelli, V. Tonchev, A. Videcoq, M. Vladimirova, Scaling and Universality of Self-Organized Patterns on Unstable Vicinal Surfaces, Phys. Rev. Lett. 88 (2002) 206103. **(DOI: 10.1103/PhysRevLett.88.206103)**

[163] J. Krug, V. Tonchev, S. Stoyanov, A. Pimpinelli, Scaling properties of step bunches induced by sublimation and related mechanisms, Phys. Rev. B 71 (2005) 045412. **(DOI: 10.1103/PhysRevB.71.045412)**

[164] R. Uecker, B. Velickov, D. Klimm, R. Bertran, M. Bernhagen, M. Rabe, M. A lbrecht, R. Fornari, D. G. Schlom, Properties of rare-earth scandate single crystals (Re=Nd-Dy), J. Cryst. Growth 310 (2008) 2649-2658. **(DOI: 10.1016/j.jcrysgro.2008.01.019)**

[165] J. H. Haeni, P. Irvin, W. Chang, R. Uecker, P. Reiche, Y. L. Li, S. Choudhury, W. Tian, M. E. Hawley, B. Craigo, A. K. Tagantsev, X. Q. Pan, S. K. Streiffer, L. Q. Chen, S. W. Kirchoefer, J. Levy, D. G. Schlom, Room-temperature ferroelectricity in strained $SrTiO_3$, Nature (London) 430 (2004) 758-761. **(DOI: 10.1038/nature02773)**

[166] J. H. Lee, Fang L, L. F. E. Vlahos, X. Ke, Y. W. Jung, L. F. Kourkoutis, J.-W. Kim, P. J. Ryan, T. Heeg, M. Roeckerath, V. Goian, M. Bernhagen, R. Uecker, P. C. Hammel, K. M. Rabe, S. Kamba, J. Schubert, J. W. Freeland, D. A. Muller, C. J. Fennie, P. Schiffer, V. Gopalan, E. J.-Halperin, D. G. Schlom, A strong ferroelectric ferromagnet created by means of spin–lattice coupling, Nature (London) 466 (2010) 954-958. **(DOI: 10.1038/nature09331)**

[167] R. Dirsyte, J. Schwarzkopf, G. Wagner, R. Fornani, J. Lienemann, M. Busch, H. Winter, Thermal-induced change in surface termination of $DyScO_3$ (110), Surf. Sci. 604 (2010) L55-L58. **(DOI: 10.1016/j.susc.2010.07.029)**

[168] J. E. Kleibeuker, G. Koster, W. Siemons, D. Dubbink, B. Kuiper, J. L. Blok, C. -H. Yang, J. Ravichandran, R. Ramesh, J. E. ten Elshof, D. H. A. Blank, G. Rijnders, Atomically Defined Rare-Earth Scandate Crystal Surfaces, Adv. Funct. Mater. 20 (2010) 3490-3496. **(DOI: 10.1002/adfm.201000889)**

[169] J. E. Kleibeuker, B. Kuiper, S. Harkema, D. H. A. Blank, G. Koster, G. Rijnders, P. Tinnemans, E. Vlieg, P. B. Rossen, W. Siemons, G. Portale, J. Ravichandran, J. M. Szepieniec, R. Ramesh, Structure of singly terminated polar $DyScO_3$ (110) surfaces, Phys. Rev. B 85 (2012) 165413. **(DOI: 10.1103/PhysRevB.85.165413)**

[170] G. F. Ruse, S. Geller, Growth of neodymium gallium oxide crystals, J. Cryst. Growth 29 (1975) 305-308. **(DOI: 10.1016/0022-0248(75)90176-1)**

[171] M. Sasaura, S. Miyazawa, Twin-free single crystal growth of $NdGaO_3$, J. Cryst. Gowth 131 (1993) 413-418. **(DOI: 10.1016/0022-0248(93)90189-4)**

[172] T. Ohnishi, K. Takahashi M. Nakamura, M. Kawasaki, M. Yoshimoto, H. Koinuma, A-site layer terminated perovskite substrate: $NdGaO_3$, Appl. Phys. Lett. 74 (1999) 2531. **(DOI: 10.1063/1.123888)**

[173] R. Gunnarssson, A. S. Kalabukhov, D. Winkler, Evaluation of recipes for obtaining single terminated perovskite oxide substrates, Surf. Sci. 603 (2009) 151-157. **(DOI: 10.1016/j.susc.2008.10.045)**

[174] M. Radovic, N. Lampis, F. Miletto, Granozio, P. Perna, Z. Ristic, M. Salluzzo, C. M. Schlepütz, U. Scotti di Uccio, Growth and characterization of stable SrO-terminated $SrTiO_3$ surfaces, Appl. Phys. Lett. 94 (2009) 022901. **(DOI: 10.1063/1.3052606)**

[175] E. Talik, M. Kruczek, H. Sakowska, Z. Ujma, M. Gała, M. Neumann, XPS characterisation of neodymium gallate wafers, J. Alloy Compd. 377 (2004) 259-267. **(DOI: 10.1016/j.jallcom.2004.01.037)**

[176] R. Dirsyte, J. Schwarzkopf, G. Wagner, J. Lienemann, M. Busch, H. Winter, R. Fornari, Surface termination of the $NdGaO_3$ (110), Appl. Surf. Sci. 255 (2009) 8685-8687. **(DOI: 10.1016/j.apsusc.2009.06.052)**

[177] A. Cavallaro, G. F. Harrington, S. J. Skinner, J. A. Kilner, Controlling the surface termination of $NdGaO_3$ (110): the role of the gas atmosphere, Nanoscale 6 (2014) 7263-7273. **(DOI: 10.1039/C4NR00632A)**

[178] V. Leca, D. H. A. Blank, G. Rijnders, Termination control of $NdGaO_3$ crystal surfaces by selective chemical etching, arXiv:1202.2256.

[179] M. Minohara, T. Tachikawa, Y. Nkanishi, Y. Hikita, L. F. Kourkoutis, J. –S. Lee, C. –C. Kao, M. Yoshita, H. Akiyama, C. Bell, H. Y. Hwang, Atomically

Engineered Metal-Insulator Transition at the $TiO_2$/$LaAlO_3$ Heterointerface, Nano. Lett. 14 (2014) 6743-6746. **(DOI: 10.1021/nl5039192)**

[180] R. J. Francis, S. C. Moss, A. J. Jacobson, X-ray truncation rod analysis of the reversible temperature-dependent [001] surface structure of $LaAlO_3$, Phys. Rev. B 64 (2001) 235425. **(DOI: 10.1103/PhysRevB.64.235425)**

[181] D. –W. Kim, D. –H. Kim, B. –S. Kang, T. W. Noh, D. R. Lee, K.-B. Lee, Roles of the first atomic layers in growth of $SrTiO_3$ films on $LaAlO_3$ substrates, Appl. Phys. Lett. 74 (1999) 2176. **(DOI: 10.1063/1.123792)**

[182] Z. L. Wang, A. J. Shapiro, Studies of $LaAlO_3$“100” surfaces using RHEED and REM. I: twins, steps and dislocations, Surf. Sci. 328 (1995) 141-158. **(DOI: 10.1016/0039-6028(95)00014-3)**

[183] J. Yao, P. B. Merrill, S. S. Perry, D. Marton, J. W. Rabalais, Thermal stimulation of the surface termination of $LaAlO_3${100}, J. Chem. Phys. 108 (1998) 1645. **(DOI: 10.1063/1.475535)**

[184] Z. Q. Liu, Z. Huang, W. M. Lu, K. Gopinadhan, X. Wang, A. Annadi, T. Venkatesan, Ariando, Atomically flat interface between a single-terminated $LaAlO_3$ substrate and $SrTiO_3$ thin film is insulating, AIP Advances 2 (2012) 012147. **(DOI: 10.1063/1.3688772)**

[185] C. H. Lanier, J. M. Rondinelli, B. Deng, R. Kilaad, K. R. Poeppelmeier, L. D. Marks, Surface Reconstruction with a Fractional Hole: $(\sqrt{5}\times\sqrt{5})R26.6^\circ$ $LaAlO_3$ (001), Phys. Rev. Lett. 98 (2007) 086102. **(DOI: 10.1103/PhysRevLett.98.086102)**

[186] A. J. H. van der Torren, S. J. van der Molen, J. Aarts, Formation of a mixed ordered termination on the surface of $LaAlO_3$ (001), Phys. Rev B 91 (2015) 245426. **(DOI: 10.1103/PhysRevB.91.245426)**

[187] K. Krishnaswamy, C. E. Dreyer, A. Janotti, C. G. Van de Walle, Structure and energetics of $LaAlO_3$ (001) surfaces, Phys. Rev. B 90 (2014) 235436. **(DOI: 10.1103/PhysRevB.90.235436)**

[188] S. Geller, E. A. Wood, Crystallographic studies of perovskite-like compounds. I. Rare earth orthoferrites and $YFeO_3$, $YCrO_3$, $YAlO_3$, Acta Cryst. 9 (1956) 563-568. **(DOI: 10.1107/S0365110X56001571)**

[189] R. Diehu, G. Brandt, Crystal structure refinement of $YAlO_3$, a promising laser material, Mater. Res. Bull. 10 (1975) 85-90. **(DOI: 10.1016/0025-5408(75)90125-7)**

[190] F. Liu, T. Makino, T. Yamaseki, K. Ueno, A. Tsukazaki, T. Fukumura, Y. Kong, M. Kawasakai, Ultrafast Time-Resolved Faraday Rotation in EuO Thin Films, Phys. Rev. Lett. 108 (2012) 257401. **(DOI: 10.1103/PhysRevLett.108.257401)**

[191] T. Inoue, T. Morimoto, S. Kaneko, Y. Horii, Y. Ohki, Crystalline structures of $YAlO_3$ single crystal at high temperatures, Proceedings of 2014 International Symposium on Electrical Insulating Materials (ISEIM), 208, (2014). **(DOI: 10.1109/ISEIM.2014.6870755)**

[192] W. Prusseit, L. A. Boatner, D. Rytz, Epitaxial $YBa_2Cu_3O_7$ growth on $KTaO_3$ (001) single crystals, Appl. Phys. Lett. 63 (1993) 3376. **(DOI: 10.1063/1.110150)**

[193] Y. Kim, A. Erbil, L. A. Boatner, Substrate dependence in the growth of epitaxial $Pb_{1-x}La_xTiO_3$ thin films, Appl. Phys. Lett. 69 (1996) 2187-2189. **(DOI: 10.1063/1.117160)**

[194] J. Sigman, D. P. Norton, H. M. Christen, P. H. Fleming, L. A. Boatner, Antiferroelectric Behavior in Symmetric $KNbO_3$/$KTaO_3$ Superlattices, Phys. Rev. Lett. 88 (2002) 097601. **(DOI: 10.1103/PhysRevLett.88.097601)**

[195] K. Zou, S. Ismail-Beigi, K. Kisslinger, X. Shen, D. Su, F. J. Walker, C. H. Ahn, $LaTiO_3$/$KTaO_3$ interfaces: A new two-dimensional electron gas system, APL Mater. 3 (2015) 036104. **(DOI: 10.1063/1.4914310)**

[196] Y. Wang, W. Tang, J. Cheng, M. Behtash, K. Yang, Creating Two-Dimensional Electron Gas in Polar/Polar Perovskite Oxide Heterostructures: First-Principles Characterization of $LaAlO_3$/$A^+B^{5+}O_3$, ACS Appl. Mater. Interfaces 8 (2016) 13659-13668. **(DOI: 10.1021/acsami.6b02399)**

[197] K. Ueno, S. Nakamura, H. Shimotani, H. T. Yuan, N. Kimura, T. Nojima, H. Aoki, Y. Iwasa, M. Kawasaki, Discovery of superconductivity in $KTaO_3$ by electrostatic carrier doping, Nat. Nano. 6 (2011) 408-412. **(DOI: 10.1038/nnano.2011.78)**

[198] Hyung-jin Bae, J. Sigman, D. P. Norton, L. A. Boatner, Surface treatment for forming unit cell steps on the (001) $KTaO_3$ substrate surface, Appl. Surf. Sci. 241 (2005) 271-278. **(DOI: 10.1016/j.apsusc.2004.05.293)**

[199] H. Nakamura, T. Kimura, Threshold electric fields controlled by surface treatments in $KTaO_3$ field-effect transistors, J. Appl. Phys. 107 (2010) 074508. **(DOI: 10.1063/1.3372712)**

[200] A. Ogawa, N. Inoue, T. Sugano, S. Adachi, K. Suzuki, N. Nakagaki, Y. Enomoto, K. Tanabe, Fabrication and characterization of high-quality (Hg,Re)$Ba_2CaCu_2O_y$ thin films on LSAT substrates, Supercond. Sci. Technol. 15 (2002) 1706. **(DOI: 10.1088/0953-2048/15/12/315)**

[201] H. Li, L. Salamanca-Riba, R. Ramesh, J. H. Scott, Ordering in (La,Sr)(Al,Ta)$O_3$ substartes, J. Mater. Res. 18 (2003) 1698. **(DOI: 10.1557/JMR.2003.0233)**

[202] M. Paradinaas, L. Garzon, F. Sánchez, R. Bachelet, D. B. Amabilino, J. Fontcuberta, C. Ocal, Tuning the local frictional and electrostatic responses of nanostructured $SrTiO_3$-surfaces by self-assembled molecular monolayers, Phys. Chem. Chem. Phys. 12 (2010) 4452-4458. **(DOI: 10.1039/B924227A)**

[203] R. Bachelet, C. Ocal, L. Garzon, J. Fontcuberta, F. Sánchez, Conducted growth of $SrRuO_3$ nanodot arrays on self-ordered $La_{0.18}Sr_{0.82}Al_{0.59}Ta_{0.41}O_3$(001) surfaces, Appl. Phys. Lett. 99 (2011) 051914. **(DOI: 10.1063/1.3622140)**

[204] H. Boschker, M. Mathews, P. Brinks, E. Houwman, A. Vailionis, G. Koster, D. H. A. Blank, G. Rijnders, Uniaxial contribution to the magnetic anisotropy of $La_{0.67}Sr_{0.33}MnO_3$ thin films induced by orthorhombic crystal structure, J. Magn. Mag. Mat. 323 (2011) 2632-2638. **(DOI: 10.1016/j.jmmm.2011.05.051)**

[205] X. C. Fan, X. M. Chen, X. Q. Liu, Structural Dependence of Microwave Dielectric Properties of $SrRAlO_4$ (R = Sm, Nd, La) Ceramics: Crystal Structure Refinement and Infrared Reflectivity Study, Chem. Mater. 20 (2008) 4092-4098. **(DOI: 10.1021/cm703273z)**

[206] J. -P. Locquet, J. Perret, J. Fompeyrine, E. Mächler, J. W. Seo, G. Van Tendeloo, Doubling the critical temperature of $La_{1.9}Sr_{0.1}CuO_4$ using epitaxial strain, Nature, 394 (1998) 453-456. **(DOI: 10.1038/28810)**

[207] A. Gozar, G. Logvenov, L. Fitting Kourkoutis, A. T. Bollinger, L. A. Giannuzzi, D. A. Muller, I. Bozovic, High-temperature interface superconductivity between metallic and insulating copper oxides, Nature 455 (2008) 782-785. **(DOI: 10.1038/nature07293)**

[208] A. E. Becerra-Toledo, L. D. Marks, Strontium oxide segregation at $SrLaAlO_4$ surfaces, Surf. Sci. 604 (2010) 1476. **(DOI: 10.1016/j.susc.2010.05.011)**

[209] A. Biswas, P. B. Rossen, J. Ravichandran, Y.-H. Chu, Y.-W. Lee, C.-H. Yang, R. Ramesh, Y. H. Jeong, Selective A- or B-site single termination on surfaces of layered oxide $SrLaAlO_4$, Appl. Phys. Lett.102 (2013) 051603. **(DOI: 10.1063/1.4790575)**

[210] A. Kalabukov, R. Gunnarsson, J. Börjesson, E. Olsson, T. Claeson, D. Winkler, Effect of oxygen vacancies in the $SrTiO_3$ substrates on the electrical properties of the $LaAlO_3/SrTiO_3$ interface, Phys. Rev. B 75 (2007) 121404(R). **(DOI: 10.1103/PhysRevB.75.121404)**

[211] M. E. Zvanut, S. Jeddy, E. Towett, G. M. Janowski, C. Brooks, D. G. Schlom, An annealing study of an oxygen vacancy related defects in $SrTiO_3$ substrates, J. Appl. Phys. 104 (2008) 064122. **(DOI: 10.1063/1.2986244)**

[212] Z. Q. Liu, C. J. Li, W. M. Lu, X. H. Huang, S. W. Zeng, X. P. Qiu, L. S. Huang, A. Annadi, J. S. Chen, J. M. D. Coey, T. Venkatesan, Ariando, Origin of the Two-Dimensional Electron Gas at $LaAlO_3/SrTiO_3$ interfaces: The Role of Oxygen Vacancies and Electronic Reconstruction, Phys. Rev. X 3 (2013) 021010. **(DOI: 10.1103/PhysRevX.3.021010)**

[213] R. D. Shannon, R. A. Oswald, J. B. Parise, B. T. Chai, P. Byszewski, A. Pajaczkowska, R. Sobolewski, Dielectric constants and crystal structures of $CaYAlO_4$, $CaNdAlO_4$, and $SrLaAlO_4$, and deviations from the oxide additivity rule, J. Solid State Chem. 98 (1992) 90-98. **(DOI: 10.1016/0022-4596(92)90073-5)**

[214] R. Uecker, R. Bertram, M. Brützam, Z. Galazka, T. M. Gesing, C. Guguschev, D. Klimm, M Klupsch, A. Kwasniewski, D. G. Schlom, Large-lattice-parameter perovskite single-crystal substrates, J. Cryst. Growth, 457 (2017) 137-142. **(DOI: 10.1016/j.jcrysgro.2016.03.014)**

[215] H. Zama, Y. Ishii, H. Yamamoto, T. Morishita, Atomically Flat MgO Single-Crystal Surface Prepared by oxygen, Jpn. J. Appl. Phys. Part 2 40 (2001) L465. **(DOI: 10.1143/JJAP.40.L465)**

[216] D. A. Pawlak, M. Ito, L. Dobrzycki, K. Wozniak, M. Oku, K. Shimamura, T. Fukuda, Structure and spectroscopic properties of (AA')(BB')$O_3$ mixed-perovskite crystals, J. Mater. Res. 20 (2005) 3329-3337. **(DOI: 10.1557/jmr.2005.0412)**

[217] Y. Z. Chen, N. Bovet, F. Trier, D. V. Christensen, F. M. Qu, N. H. Andersen, T. Kasama, W. Zhang, R. Giraud, J. Dufouleur, T.S. Jespersen, J. R. Sun, A. Smith, J. Nygård, L. Lu, B. Bu¨chner, B.G. Shen, S. Linderoth, N. Pryds, A high-mobility two-dimensional electron gas at the spinel/perovskite interface of γ-$Al_2O_3$/$SrTiO_3$, Nat. Commun. 4 (2013) 1371. **(DOI: 10.1038/ncomms2394)**

[218] F. J. Wong, S. Ramanathan, Nonisostructural complex oxide heteroepitaxy, J. Vac. Sci. Technol. A 32 (2014) 040801. **(DOI: 10.1116/1.4879695)**

[219] J. A. Moyer, R Gao, P. Schiffer, L. W. Martin, Epitaxial growth of highly-crystalline spinel ferrite thin films on perovskite substrates for all-oxide devices, Sci. Rep. 5 (2015) 10363. **(DOI: 10.1038/srep10363)**

[220] M. O'Sullivan, J. Hadermann, M. S. Dyer, S. Turner, J. Alaria, T. D. Manning, A. M. Abakumov, J. B. Claridge, M. J. Rosseinsky, Interface control by chemical and dimensional matching in an oxide heterostructure, Nat. Chem. 8 (2016) 347-353. **(DOI: 10.1038/nchem.2441)**

[221] D. Lu, D. J. Baek, S. S. Hong, L. F. Kourkotis, Y. Hikita, H. Y. Hwang, Synthesis of freestanding single-crystal perovskite films and heterostructures by etching of sacrificial water-soluble layers, Nat. Mater. 15 (2016) 1255-1260. **(DOI: 10.1038/nmat4749)**

[222] M. Yoshimoto, T. Maeda, T. Ohnishi, H. Koinuma, O. Ishiyama, M. Shinohara, M. Kubo, R. Miura, A. Miyamoto, Atomic-scale formation of ultrasmooth surfaces on sapphire substrates for high-quality thin-film fabrication, Appl. Phys. Lett. 67 (1995) 2615. **(DOI: 10.1063/1.114313)**

[223] Y. Wang, S. Lee, P. Vilmercati, H. N. Lee, H. H. Weitering, P. C. Snijders, Atomically flat reconstructed rutile $TiO_2$(001) surfaces for oxide film growth, Appl. Phys. Lett. 108 (2016) 091604. **(DOI: 10.1063/1.4942967)**

[224] M. Khalid, A. Setzer, M. Ziese, P. Esquinazi, D. Spemann, A. Pöppl, E. Goering, Ubiquity of ferromagnetic signals in common diamagnetic oxide crystals, Phys. Rev. B 81 (2010) 214414. **(DOI: 10.1103/PhysRevB.81.214414)**

[225] D. A. Crandles, B. DesRoches, F. S. Razavi, A search for defect related ferromagnetism in $SrTiO_3$, J. Appl. Phys. 108 (2010) 053908. **(DOI: 10.1063/1.3481344)**

[226] S. M. M. Yee, D. A. Crandles, L.V. Goncharova, Ferromagnetism on the unpolished surfaces of single crystal metal oxide substrates, J. Appl. Phys. 110 (2011) 033906. **(DOI: 10.1063/1.3611034)**

[227] L. Bjaalie, B. Himmetoglu, L. Weston, A. Janotti, C. G. Van de Walle, Oxide interfaces for novel electronic applications, New J. Phys. 16 (2014) 025005. **(DOI: 10.1088/1367-2630/16/2/025005)**

[228] M. Lorentz, M. S. Ramachandra Rao, T. Venkatesan, E. Fortunato, P. Barquinha, R. Branquinho, D. Salgueiro, R. Martins, E. Carlos, A. Liu, F. K. Shan, M. Grundmann, H. Boschker, J. Mukherjee, M. Priyadarshini, N. DasGupta, D. J. Rogers, F. H. Teherani, E. V. Sandana, P. Bove, K. Rietwyk, A. Zaban, A. Veziridis, A. Weidenkaff, M. Muralidhar, M. Murakami, S. Abel, J. Fompeyrine, J. Zuniga-Perez, R. Ramesh, N. A. Spaldin, S. Ostanin, V. Borisov, I. Mertig, V. Lazenka, G. Srinivasan, W. Prellier, M. Uchida, M. Kawasaki, R. Pentcheva, P. Getenwart, F. Miletto Granozio, J. Fontcuberta, N. Pryds, The 2016 oxide electronic materials and oxide interfaces roadmap, J. Phys. D: Appl. Phys. 49 (2016) 433001. **(DOI: 10.1088/0022-3727/49/43/433001)**

[229] S. Lee, J. Jiang, Y. Zhang, C. W. Bark, J. D. Weiss, C. Tarantini, C. T. Nelson, H. W. Jang, C. M. Folkman, S. H. Baek, A. Polyanskii, D. Abraimov, A. Yamamoto, J. W. Park, X. Q. Pan, E. E. Hellstrom, D. C. Larbalestier, C. B. Eom, Template engineering of Co-doped $BaFe_2As_2$ single-crystal thin films, Nat. Mat. 9 (2010) 397-402. **(DOI: 10.1038/nmat2721)**

[230] J. Chakhalian, A. J. Millis, J. Rondinelli, Whither the oxide interface, Nat. Mater. 11 (2012) 92-94. **(DOI: 10.1038/nmat3225)**